\documentclass[preprint,amsmath,amssymb]{revtex4}
\usepackage{dcolumn}
\usepackage{bm}
\usepackage{graphicx}
\usepackage{subfigure}
\usepackage{color}

\newcommand{\be}{\begin{eqnarray}}
\newcommand{\ee}{\end{eqnarray}}

\newlength{\fighskip} \fighskip=2pt
\newlength{\figvskip} \figvskip=3pt
\newcommand*{\figbox}[2]{{
\def\figscale{#1}
\def\arraystretch{0.8}
\arraycolsep=0pt
\begin{array}{c}
\vbox{\vskip\figscale\figvskip
\hbox{\hskip\figscale\fighskip
\includegraphics[scale=\figscale]{#2}}}
\end{array}}}

\begin{document}

\title{On the fidelity of information retrieval in the black hole final state model with scrambling interactions}

\author{Ran Li$^{1}$}
\thanks{liran@qfnu.edu.cn}

\author{Jin Wang$^{2}$}
\thanks{Corresponding author: jin.wang.1@stonybrook.edu}

\affiliation{
 $^1$ Department of Physics, Qufu Normal University, Qufu, Shandong 273165, China\\
 $^2$ Department of Chemistry, and Department of Physics and Astronomy, The State University of New York at Stony Brook, Stony Brook, NY 11794, USA}

\begin{abstract}
We study the fidelity of information retrieval in the black hole final state model by taking into account the interactions between the collapsing matter and the infalling Hawking radiation inside the event horizon. By utilizing a scrambling unitary operator to model these interactions, our direct calculations suggest that the information is almost lost during the process of black hole evaporation. We then proceed to employ the Yoshida-Kitaev decoding strategy for information retrieval. Although we observe a finite improvement in decoding fidelity, it does not reach unity. This indicates that the issue of unitarity in black hole evaporation may not be fully resolved within the framework of the final state model.
\end{abstract} 

\maketitle

\section{Introduction} 
\label{sec:intro}

Understanding how information can be recovered from evaporating black holes is a fundamental challenge in the realm of quantum gravity. This issue is intimately related to the unitarity of the black hole evolution \cite{Hawking:1976ra}. Pioneering work by Page \cite{Page:1993df,Page:1993wv} demonstrated that if black hole evaporation process is indeed unitary, the entanglement entropy of Hawking radiation follows a characteristic curve now known as the Page curve. Especially, subsequent investigations \cite{Hayden:2007cs} from the perspective of quantum information theory indicated that in principle the information initially deposited in the black hole can be revealed in the Hawking radiation. In a notable contribution \cite{Yoshida:2017non}, Yoshida and Kitaev (YK) proposed an explicit strategy for decoding the quantum information from the Hawking radiation.

In a very interesting paper \cite{Horowitz:2003he}, Horowitz and Maldacena (HM) proposed a solution to the unitarity problem of black hole evaporation by imposing a final state boundary condition at the black hole singularity. However, this model simplifies the dynamics by neglecting the interaction between the collapsing matter and the infalling Hawking radiation. In a subsequent study \cite{Gottesman:2003up}, Gottesman and Preskill (GP) argued that incorporating these interactions would lead to a violation of unitarity in black hole evaporation. It is also shown that in the worst case, the information is completely lost. This argument was challenged by Lloyd in \cite{Lloyd:2004wn}, where the averaged fidelity is calculated by considering a random final state. He concluded that the quantum information contained in the collapsing matter escapes from the black hole with fidelity exponentially close to $1$. Later, it is shown that if employing the Schmidt form of the black hole final state to calculate the fidelity, the unitary transformation that is introduced to match the basis of the collapsing matter state and that of the outgoing Hawking radiation state depends on the initial state of the matter \cite{Lee:2020aft}. Furthermore, if considering the generic final state without a Schmidt form, it is shown that information will almost certainly be lost because the fidelity will approach zero when the degrees of freedom of the black hole are sufficiently large.

In the present work, we will consider the final state model by taking the interactions between the collapsing matter and the infalling radiation into account. These interactions can be well represented by an unitary scrambling operator $U$ for simplicity. At the black hole singularity, the state of the matter $M$ and the interior radiation $R_{in}$ is post-selected onto a fixed state $|\psi\rangle_f$. We will apply the YK decoding strategy to such a model. The explicit calculations show an imperfect decoding of Hawking radiation for such a model. This means that when considering the complicated interactions, the unitarity problem of the black hole evaporation may not be fully resolved.

\section{Black hole final state model}

\begin{figure}
        \centering
        \includegraphics[width=0.3\textwidth]{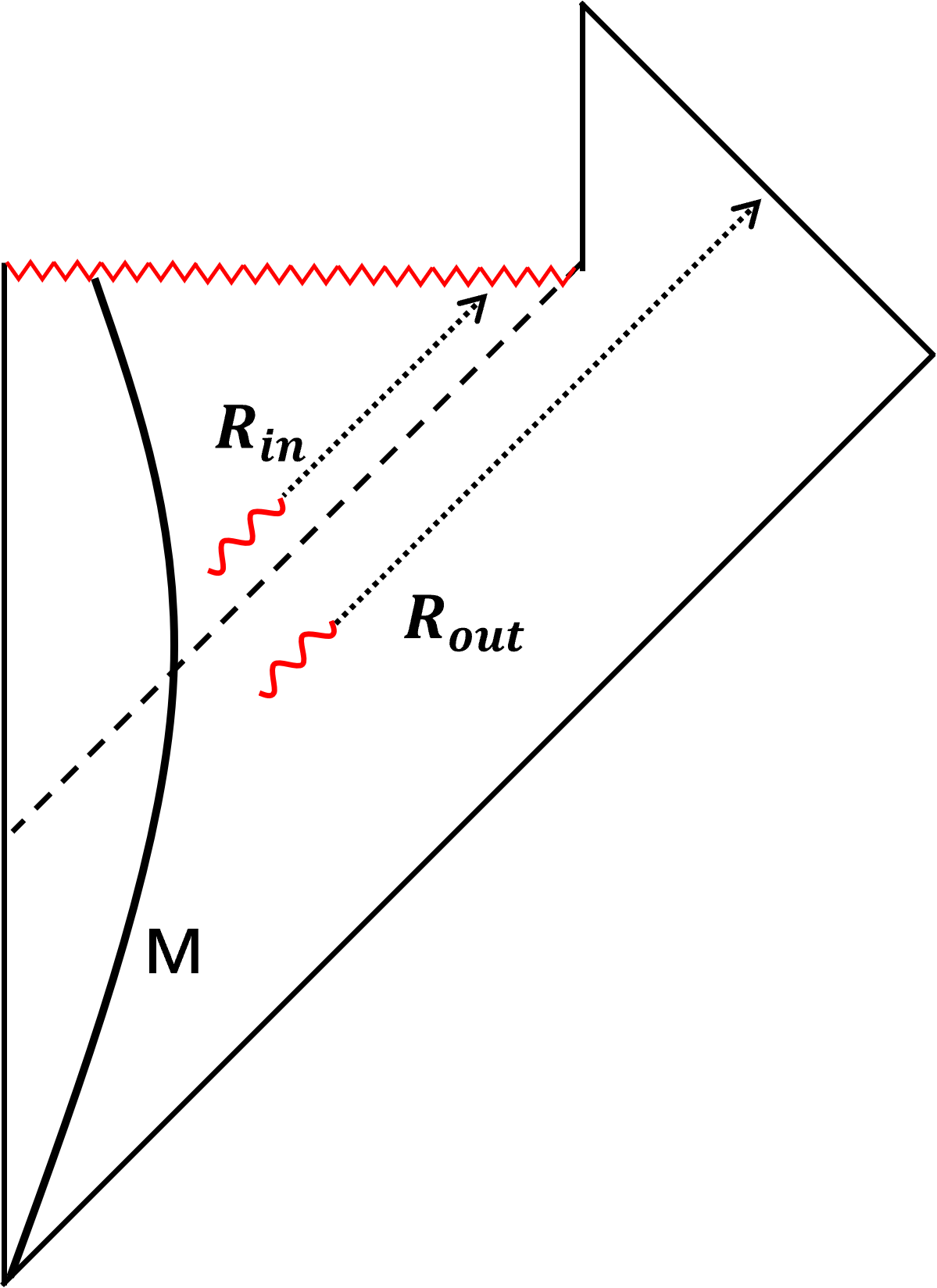}
        \caption{The Penrose diagram of black hole forming and evaporating process. The outside Hawking radiation $R_{out}$ are maximally entangled with the infalling Hawking partner modes $R_{in}$. Here, $M$ represents the matter that collapses to form black hole.}\label{Penrose_diagram}
\end{figure}

In this section, let us review the essential aspects of the black hole final state model. The Penrose diagram that describes the forming and the evaporating process of a black hole is briefly shown in Figure \ref{Penrose_diagram}. It is well known that from effective field calculations, the vacuum fluctuations evolve into the Unruh state in $\mathcal{H}_{R_{in}}\otimes \mathcal{H}_{R_{out}}$, where $\mathcal{H}_{R_{in}}$ and $\mathcal{H}_{R_{out}}$ represent the Hilbert spaces of the infalling radiation system and the outside radiation system. The initial matter state $|\psi\rangle_M$ evolves into a state in $\mathcal{H}_M\otimes \mathcal{H}_{R_{in}}\otimes \mathcal{H}_{R_{out}}$. In the micro-Canonical ensemble, considering that all states in $\mathcal{H}_{R_{out}}$ have the fixed energy equal to the black hole mass, the Unruh state in $\mathcal{H}_{R_{in}}\otimes \mathcal{H}_{R_{out}}$ is just the maximally entangled EPR state, which can be explicitly expressed as
\begin{eqnarray}
    |\Psi_{\textrm{Unruh}}\rangle_{R_{in}\otimes R_{out}}&=&|\textrm{EPR}\rangle_{R_{in}\otimes R_{out}}=\frac{1}{\sqrt{N}}\sum_{i}|i\rangle_{R_{in}} \otimes |i\rangle_{R_{out}}\;,
\end{eqnarray}
where $N=e^{S}$ is the Hilbert space dimension of black hole with $S$ being its thermodynamic entropy, and $\{|i\rangle_{R_{in}}\}$ and $\{|i\rangle_{R_{out}}\}$ are the orthonormal bases for $\mathcal{H}_{R_{in}}$ and $\mathcal{H}_{R_{out}}$ respectively.  

The effective field calculations suggest that the process of the formation and evaporation of a black hole is not unitary and the initial pure state of the collapsing matter evolves into a mixed state of radiation when the black hole disappears. To resolve this issue, HM proposed a mild modification that at the black hole singularity one needs to impose a unique final state boundary condition. 

Consider the state  
\begin{eqnarray}
    |\psi\rangle_{M\otimes R_{in} \otimes R_{out}}=|\psi\rangle_M \otimes |\Psi_{\textrm{Unruh}}\rangle_{R_{in} \otimes R_{out}}\;.
\end{eqnarray}
At the singularity, the final state boundary condition requires the quantum state of $\mathcal{H}_M\otimes \mathcal{H}_{R_{in}}$ at the singularity to post-select onto the maximally entangled state
\begin{eqnarray}
    \langle BH|=\frac{1}{\sqrt{N}}\sum_{i}~_M\langle i|\otimes~_{R_{out}}\langle i|\;.
\end{eqnarray}

The outcomes of all measurements made by an outside observer  on the outside radiation $R_{out}$ using the measurement rules in post-selected quantum mechanic are consistent with treating it as being in the pure state 
\begin{eqnarray}
      |\tilde{\psi}\rangle=N \langle BH|\psi\rangle_M \otimes |\Psi_{\textrm{Unruh}}\rangle_{R_{in} \otimes R_{out}}\;,
\end{eqnarray}
by using the ordinary quantum mechanics. This is equivalent to the unitary evolution from $|\psi\rangle_M$ to $|\tilde{\psi}\rangle$. In the final state model, the information propagates from past infinity to the black hole singularity. Instead of being absorbed at the singularity, it undergoes reflection, traveling backward in time from the singularity to the preparation of the Unruh state. From there, it undergoes another reflection and propagates to future infinity. The whole process creates an information propagation channel analogous to the quantum teleportation protocol \cite{Bennett:1992tv}.

Such a simple proposal was claimed to resolve the contradiction in the effective field calculations. Gottesman and Preskill considered the black hole final state model with the interactions between the collapsing matter and the inside Hawking partner modes \cite{Gottesman:2003up}. These interactions may break down the entanglement of the final state and result in the information loss during the black hole evaporating process. Lloyd investigated the HM proposal with the random final state and showed that information escapes with high average fidelity $\approx\left(\frac{8}{3\pi}\right)^2$. In principle, the random final state model studied by Lloyd is equivalent to the GP refinement of black hole final state model with the scrambling interactions. Therefore, he argued that the final state projection still preserves the majority of the information. Recently, it is pointed out in \cite{Lee:2020aft} that the utilizing of Schmidt form for the random final state leads to that the unitary transformation introduced to match the basis of the collapsing matter state and that of the outgoing Hawking radiation state depends on the initial state of the matter. It is also argued that information will almost certainly be lost in the random final state model.

\section{Average fidelity of black hole final model with scrambling interactions} 

 The final state model with the scrambling interactions \cite{Lloyd:2013bza} is described as follows. At the initial time, the matter system $M$ in the quantum state $|\psi\rangle_M$ collapsed to form a black hole. Due to Hawking effect, the black hole evaporated completely and became radiation. We consider the black hole final state model by taking the scrambling interactions between the collapsing matter system and the infalling radiation system into account. The interaction takes place in the black hole interior. After scrambling, the collapsing matter system and the infalling Hawking radiation are projected into a fixed pure state $|\psi\rangle_f$. The outside radiation system is in the state $|\Phi\rangle$. The whole process can be properly represented by the following graph
\begin{eqnarray}\label{Phi_state}
    |\Phi\rangle &=& N \langle \psi|_f \left(U_{M,R_{in}}\otimes I_{R_{out}}\right) |\psi\rangle_M \otimes |\textrm{EPR}\rangle_{R_{in},R_{out}}\nonumber\\
    &=& N\quad\left( \figbox{0.3}{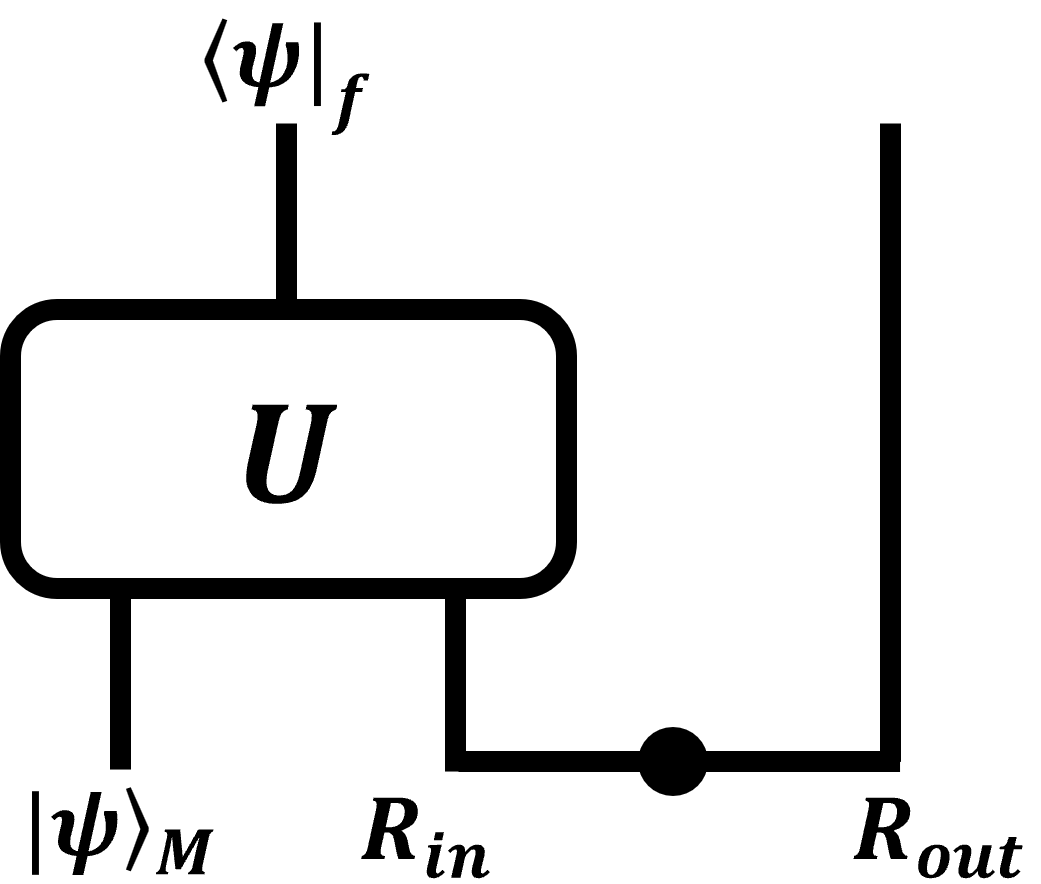}\right)\;,
\end{eqnarray}
where $N=e^S$, with $S$ the entropy of the black hole, is introduced to preserve the normalization condition, which will be checked in the following. In this graphical representation, $\figbox{0.25}{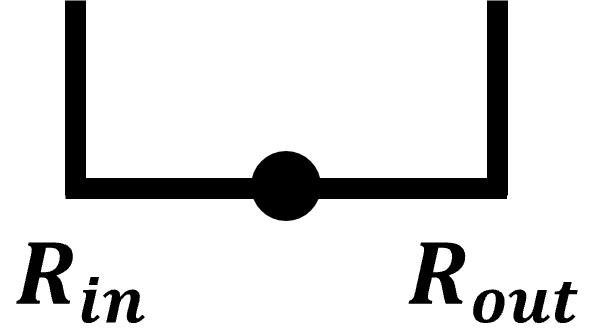}$ represents the $\textrm{EPR}$ state of the subsystems $R_{in}$ and $R_{out}$, with the black dot standing for the normalization factor $\frac{1}{\sqrt{N}}$.

Rigorously speaking, the state $|\Phi\rangle$ is not a pure state due to the final state projection. Therefore, the state can be properly described by the density matrix 
\begin{eqnarray}
    \rho_{\Phi} =  N^2 \quad\left( \figbox{0.3}{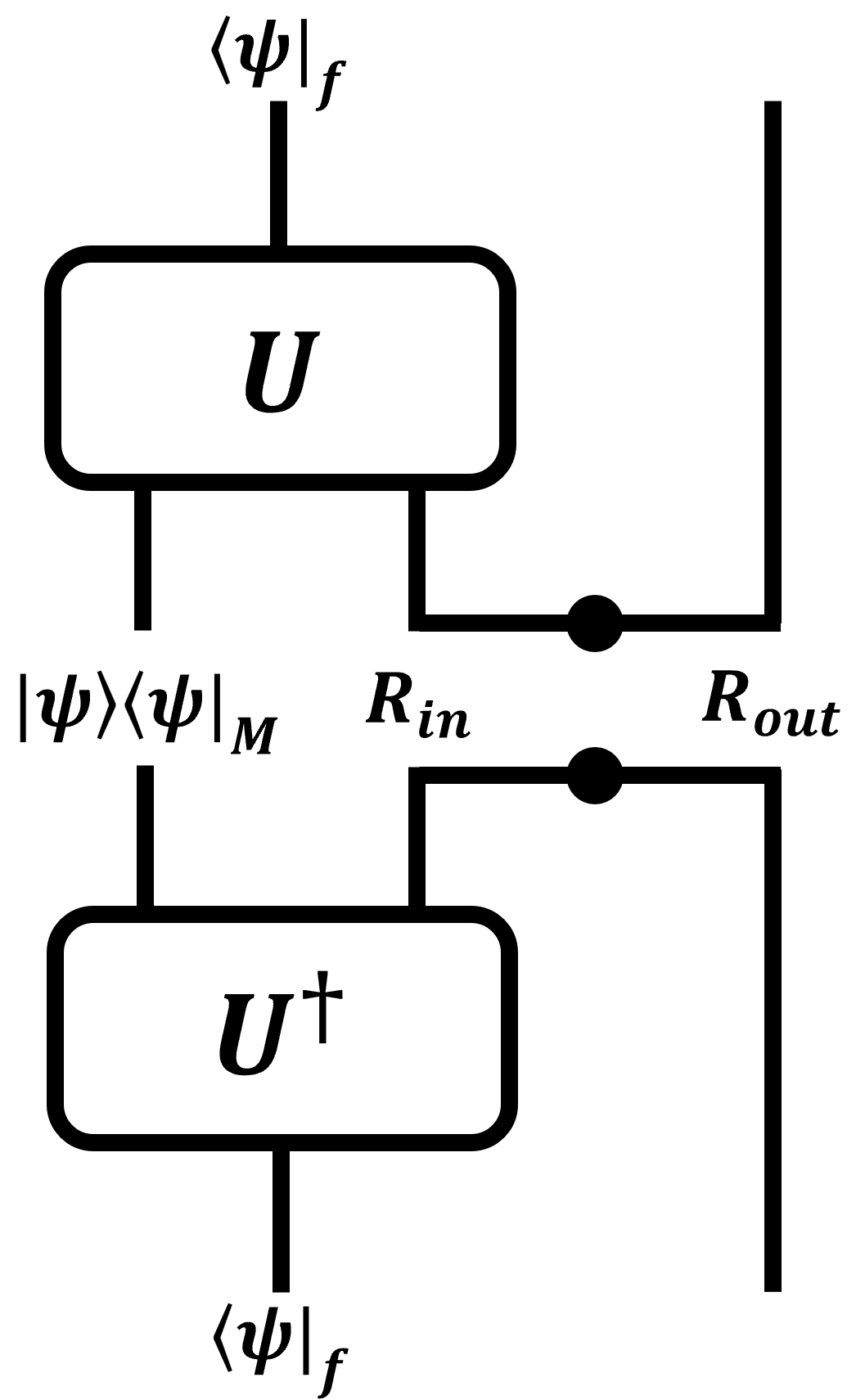}\right)\;.
\end{eqnarray}
The graph depicts the evolution of the system initially prepared in the state $|\psi\rangle_M\langle \psi| \otimes |\textrm{EPR}\rangle_{R_{in},R_{out}}\langle \textrm{EPR}|$. The system evolves forward in time through the application of a scrambling operator $U$, and backwards in time through the conjugate operator $U^\dagger$. Finally, the evolved state is projected onto the black hole final state $|\psi\rangle_f\langle \psi|$.

Now we want to check that the density matrix is normalized under the Haar average. The following Haar integral formula is useful \cite{Kim:2022pfp,Li:2024tcm}
\begin{eqnarray}
    \int dU U_{ij} U^\dagger_{j'i'}&=&\frac{1}{d}\delta_{ii'}\delta_{jj'} \;.\label{2U_integral}
\end{eqnarray}
It is convenient to represent the integral formula in the graphical form 
\begin{eqnarray}
    \int dU \left(\figbox{0.3}{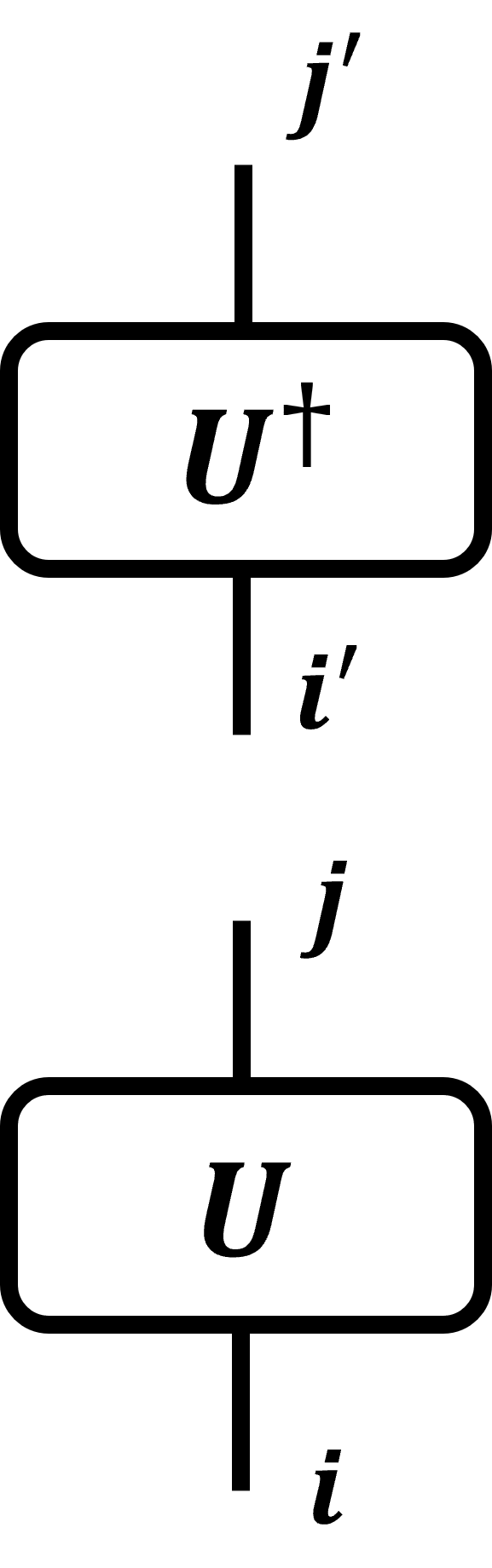}\right)&=&\frac{1}{d}\left(\figbox{0.3}{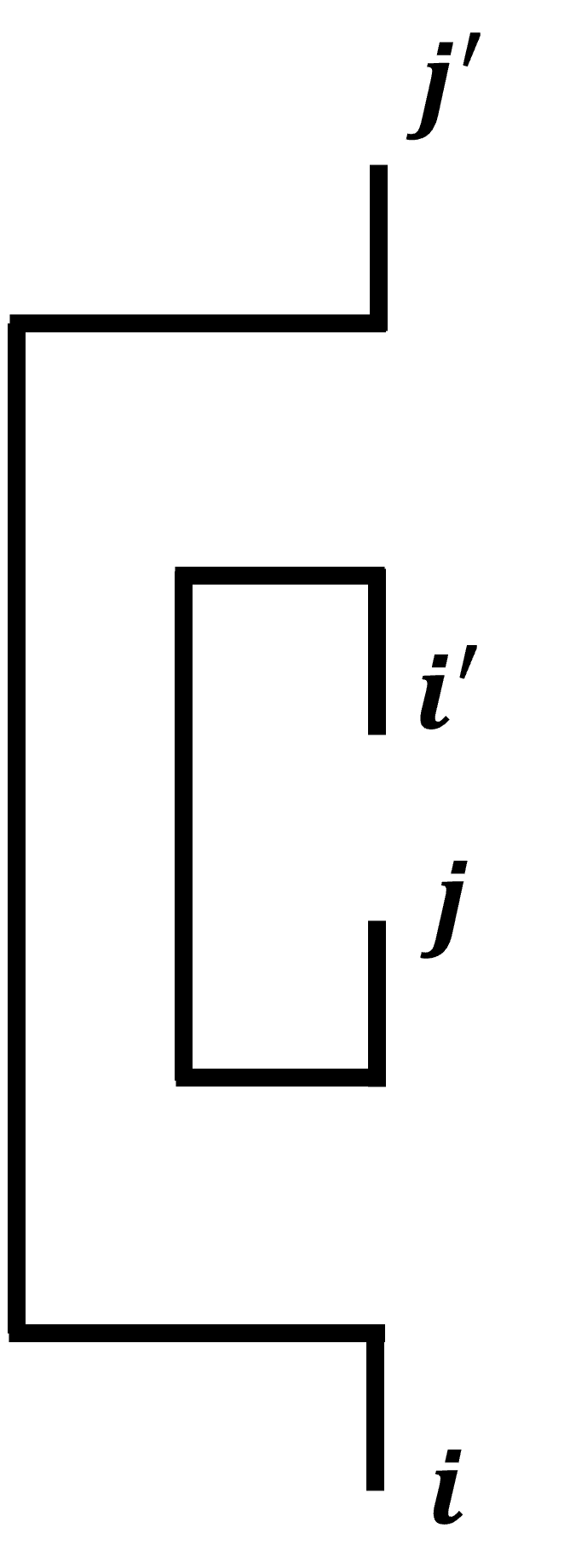}\right)\;.\label{2U_graph}
\end{eqnarray}
The density matrix $\rho_{\Phi}$ satisfies the normalization condition under the Haar average, which can be explicitly shown as
\begin{eqnarray}\label{Tr_rho_phi_graph}
  \int dU \textrm{Tr}\rho_\Phi=N\int dU \left(  \figbox{0.3}{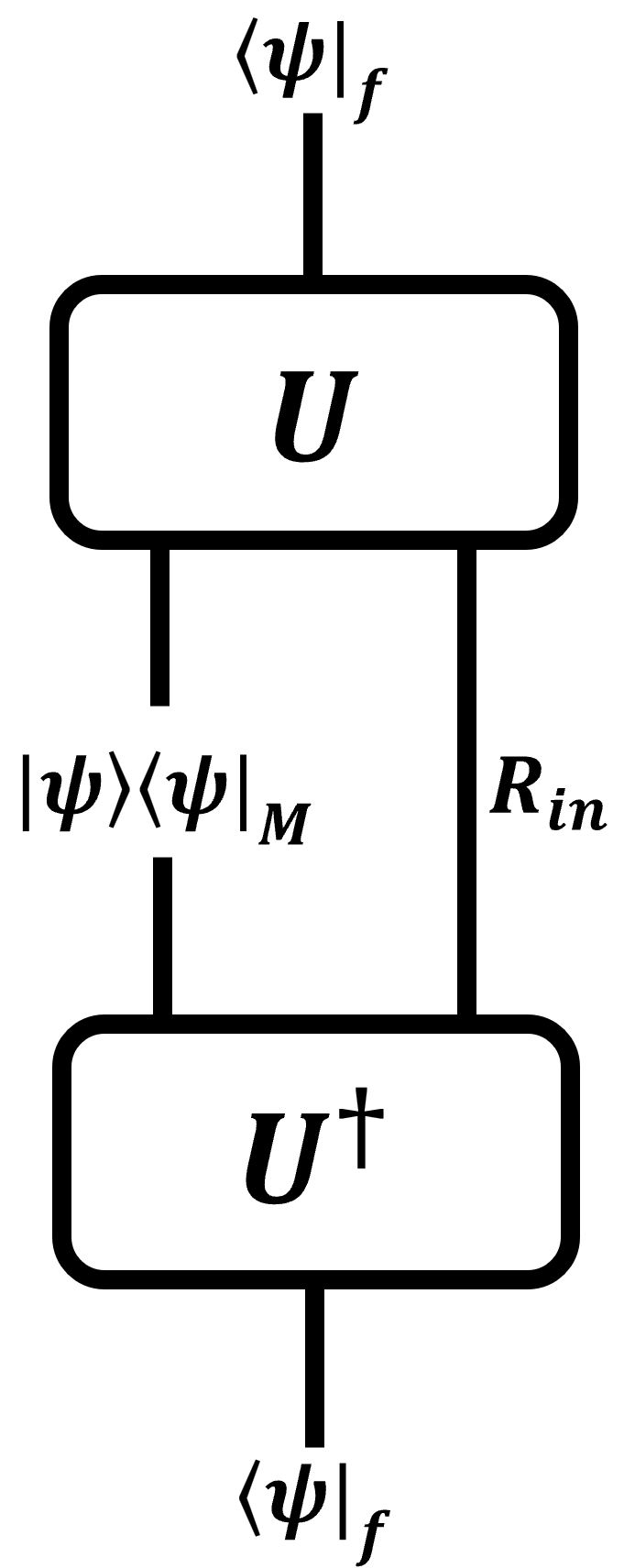} \right)=\frac{1}{N} \left(  \figbox{0.3}{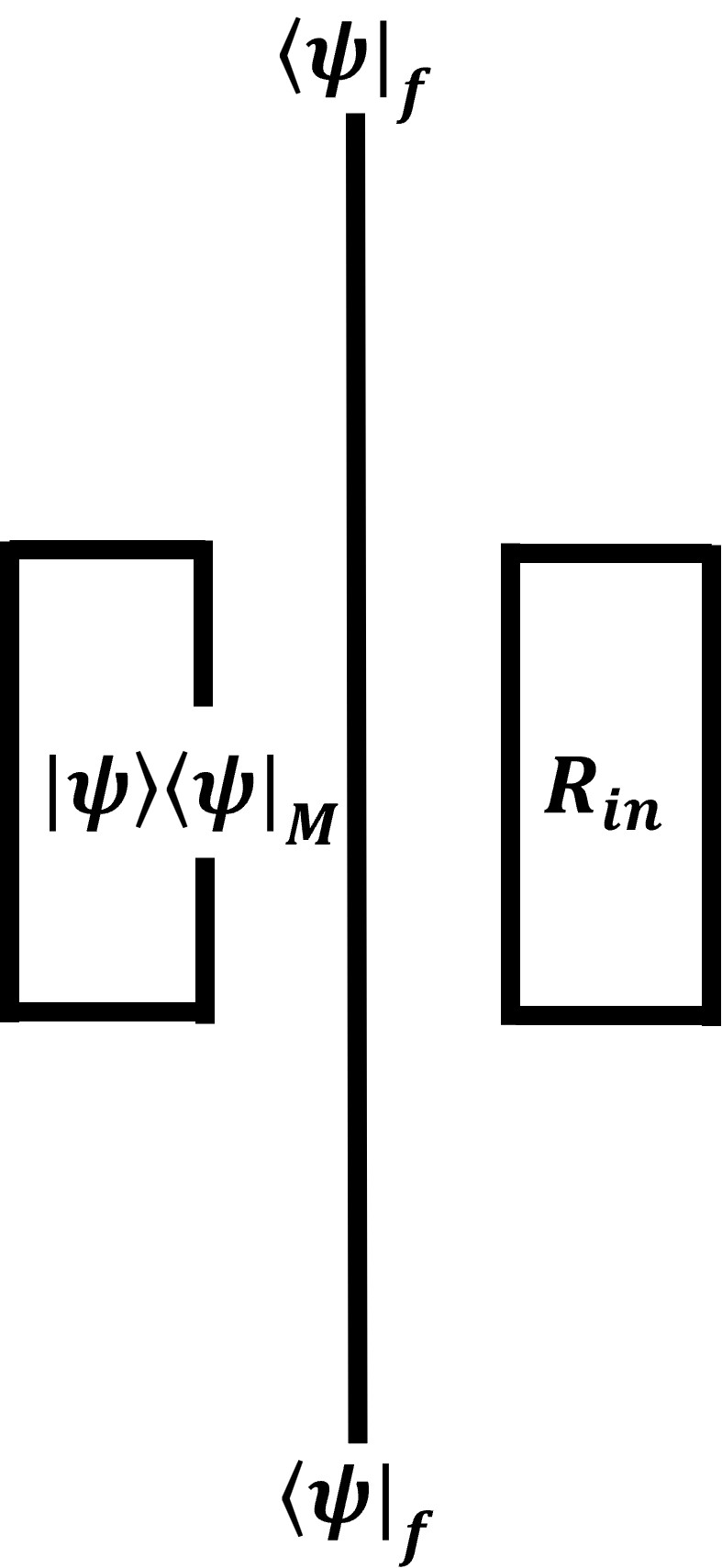} \right)=1\;.
\end{eqnarray}
In the above procedure, we used the fact that $d=N^2$ being the dimension of the state space $\mathcal{H}_{M}\otimes\mathcal{H}_{R_{in}}$ and a loop of the subsystem $R_{out}$ contributes a factor $N$ of the dimension of the Hilbert space $\mathcal{H}_{R_{out}}$. 

The fidelity between the initial collapsing matter state $|\psi\rangle_{M} \langle\psi|$ and the final radiation state $\rho_{\Phi}$ is defined as
\begin{eqnarray}
    f=\textrm{Tr} \left(|\psi\rangle_{M} \langle\psi| \rho_{\Phi} \right)
    =N^2 \quad\left( \figbox{0.3}{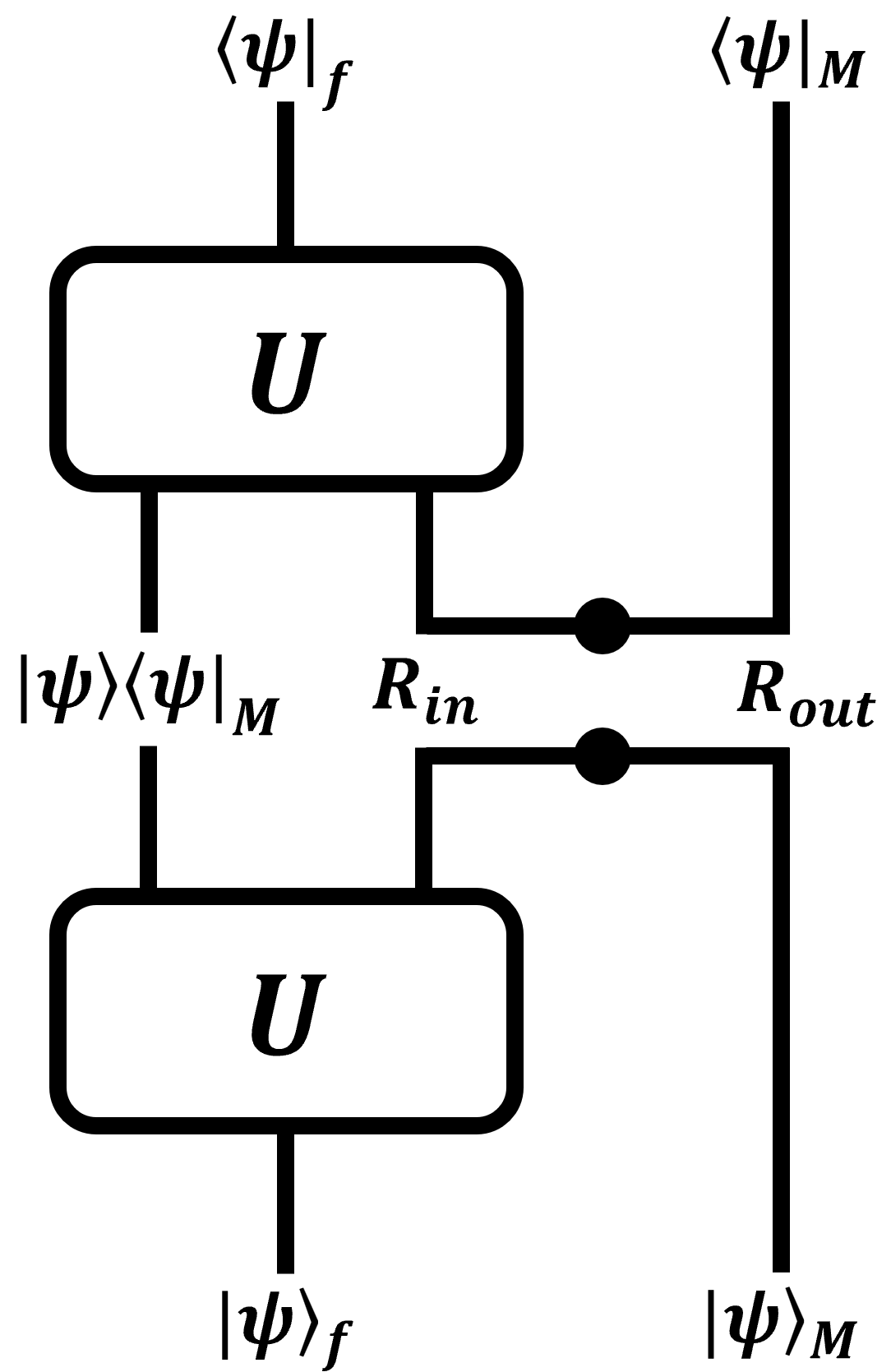}\right)=N \quad\left( \figbox{0.3}{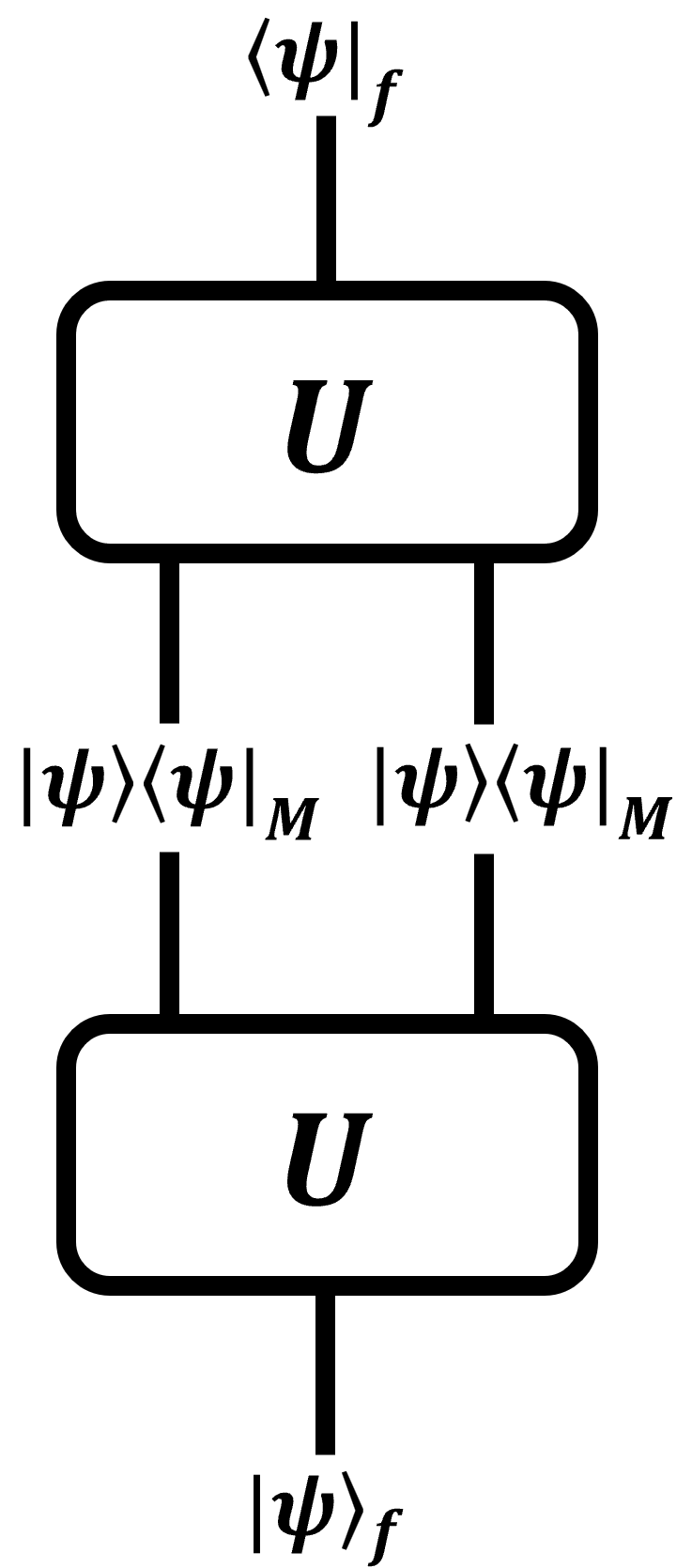}\right)\;.
\end{eqnarray}
In the last step, we used the fact that $\mathcal{H}_{in}$ and $\mathcal{H}_{out}$ are isometric. By applying the graphical representation of Haar average given in Eq.\eqref{2U_graph}, the averaged fidelity can be calculated as follows 
\begin{eqnarray}     
     \overline{f}=\int dU f=\frac{1}{N} \left( \figbox{0.3}{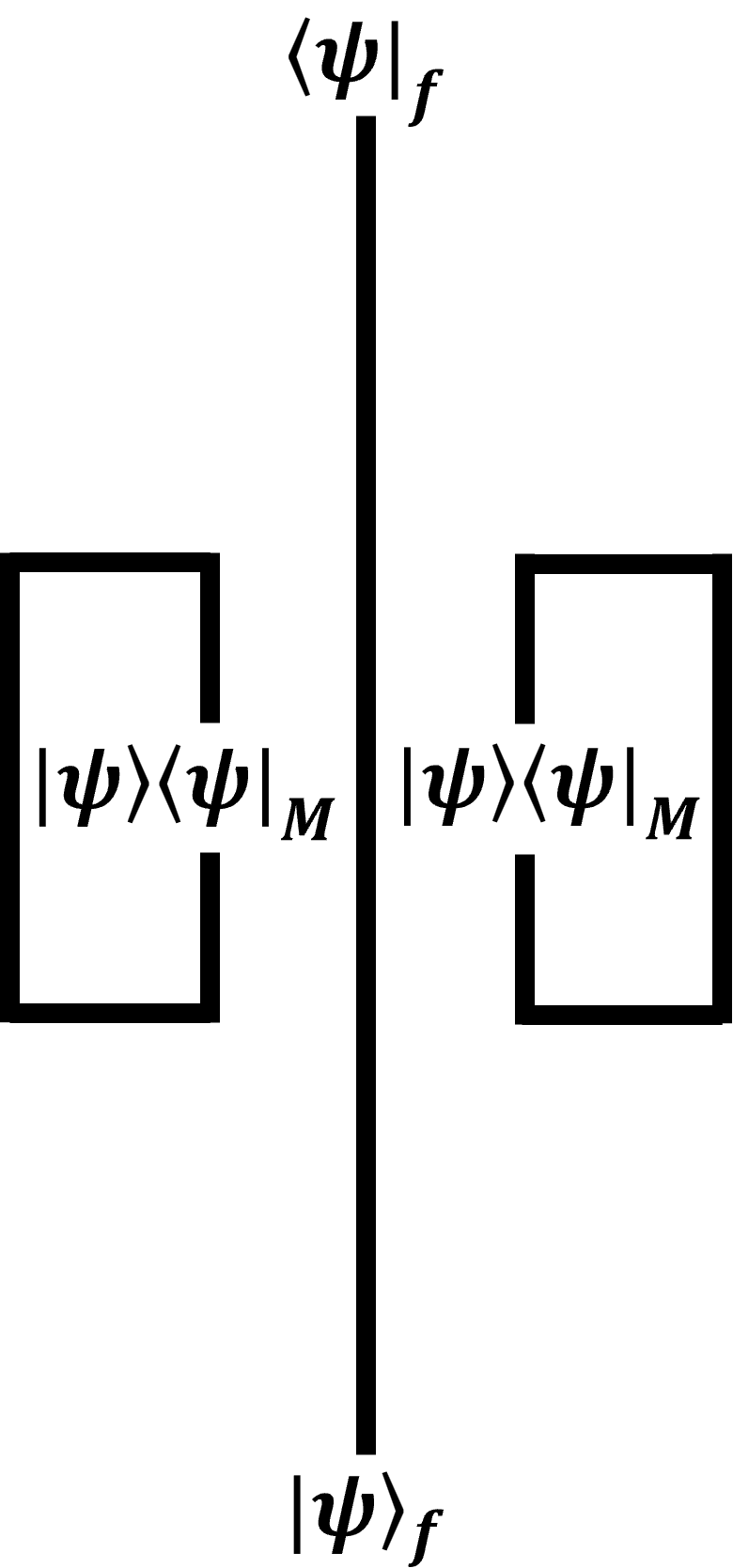}\right)=\frac{1}{N}\;.
\end{eqnarray}
This result is consistent with that obtained in \cite{Lee:2020aft}. When $N$ is sufficiently large, the fidelity is nearly zero. It means that if selecting the generic random final state or taking the interactions between the collapsing matter and the infalling radiation into account, the information is almost lost in black hole evaporation process. 

\section{Information retrieval using the Yoshida-Kitaev decoding strategy}

In this section, we apply the Yoshida-Kitaev probabilistic decoding strategy \cite{Yoshida:2017non,Yoshida:2018vly} to try to recover the initial information contained in the collapsing matter system.

Given the quantum state $|\Phi\rangle$ as graphically represented in Eq.\eqref{Phi_state}, the Yoshida-Kitaev decoding protocol proceeds as follows:

\begin{itemize}
    \item[1.] Prepare an EPR state, denoted as $|\textrm{EPR}\rangle_{\overline{D}D}$. 
    
    \item[2.] Apply the operator $U^*$ on the subsystems $R_{out}$ and $\overline{D}$. We denote the resultant state as the state $|\Psi_{in}\rangle$. It can be graphically represented as 
    \begin{eqnarray}
        |\Psi_{in}\rangle=N \quad \figbox{0.3}{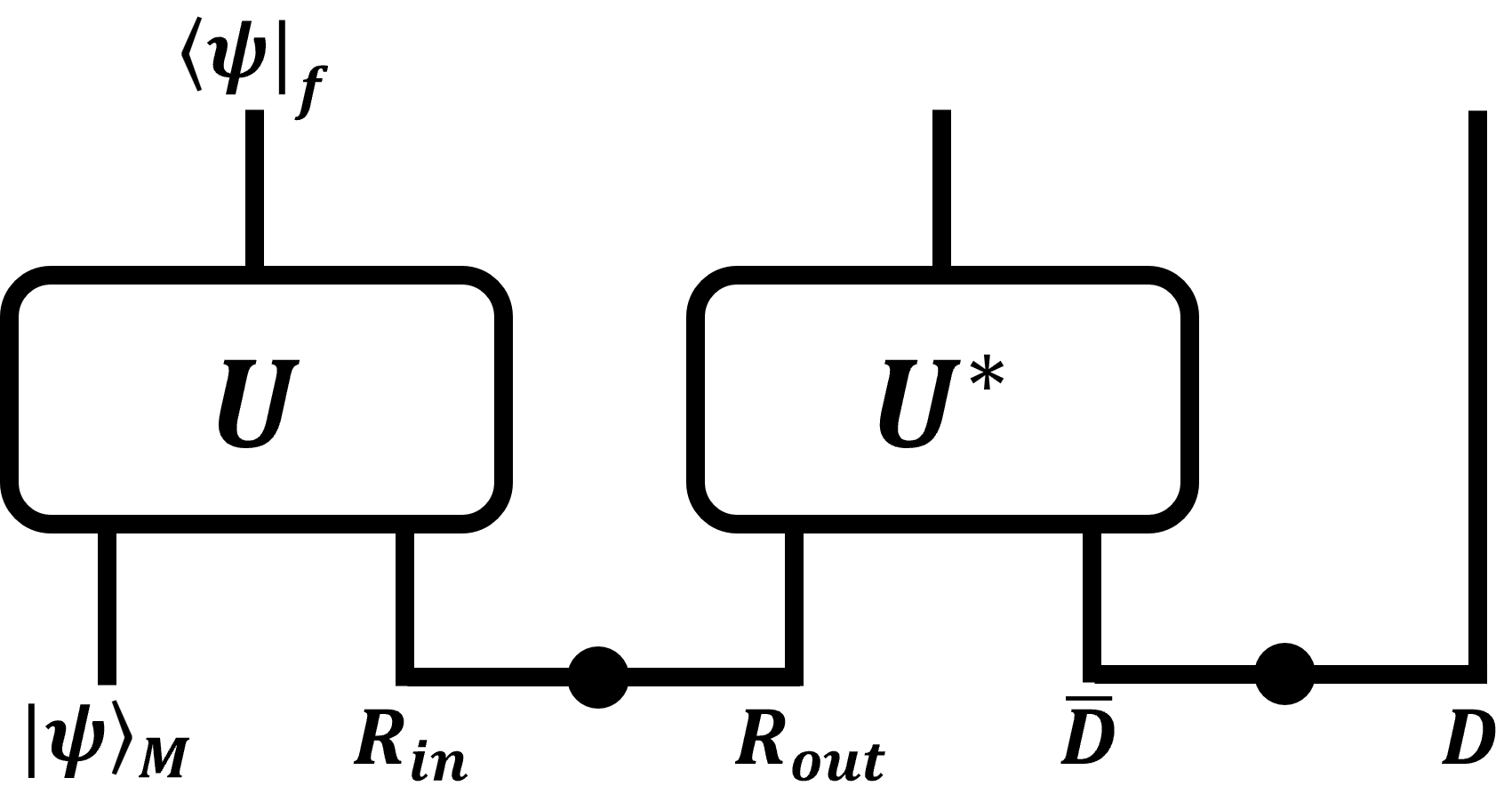}\;.
    \end{eqnarray}

    \item[3.] Do the projective measurement on the output state of the operator $U^*$. If the outcome is $|\psi\rangle_f$, it means the successful decoding of of the initial information contained in the matter system $M$. We denote the resultant state as $|\Psi_{out}\rangle$, which can be graphically represented as
    \begin{eqnarray}\label{psi_out}
        |\Psi_{out}\rangle&=&\frac{1}{\sqrt{P}} ~_{f}\langle \psi|\Psi_{in}\rangle\nonumber\\
        &=&\frac{N}{\sqrt{P}} \quad \figbox{0.3}{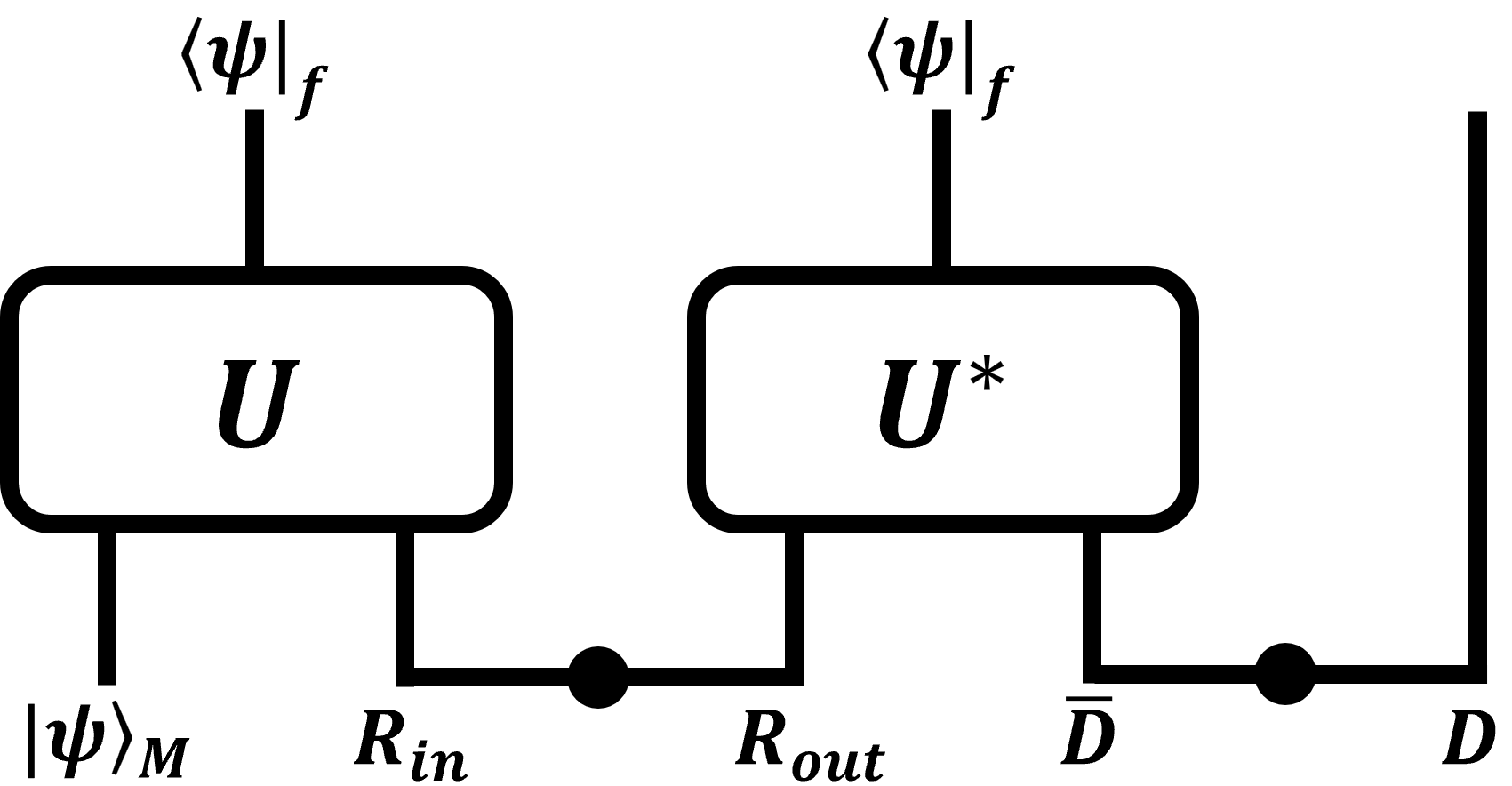}\;.
    \end{eqnarray}
The prefactor $\frac{1}{\sqrt{P}}$ is introduced to preserve the normalization of $|\Psi_{out}\rangle$. It is actually the projecting probability onto the state $|\psi\rangle_f$.  
\end{itemize}

Notice that 
\begin{eqnarray}
   && \textrm{Tr}\left[|\Psi_{in}\rangle\langle\Psi_{in}|\right]\nonumber\\&&=N \left( \figbox{0.3}{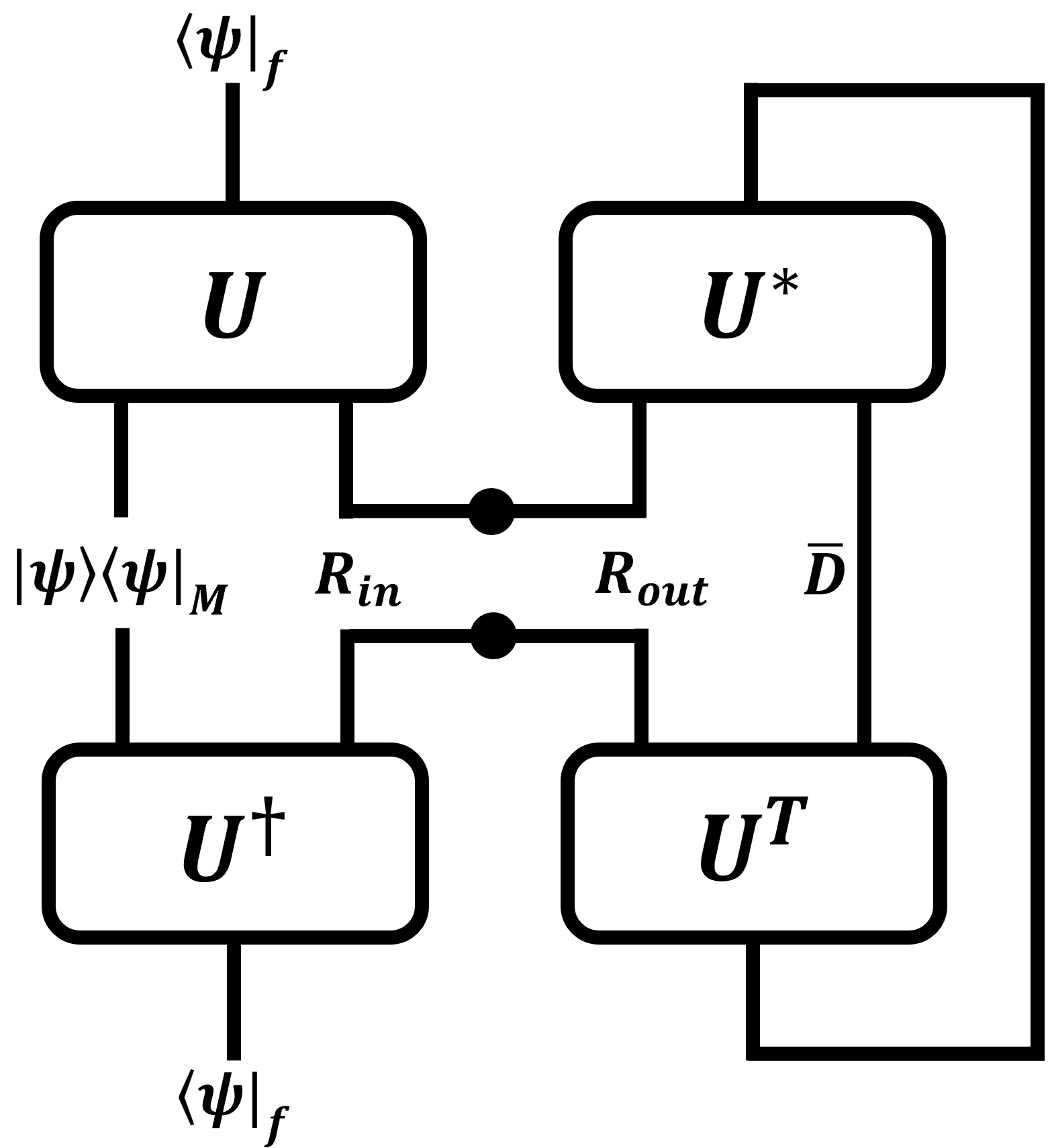}\right)
    =N \left(\figbox{0.3}{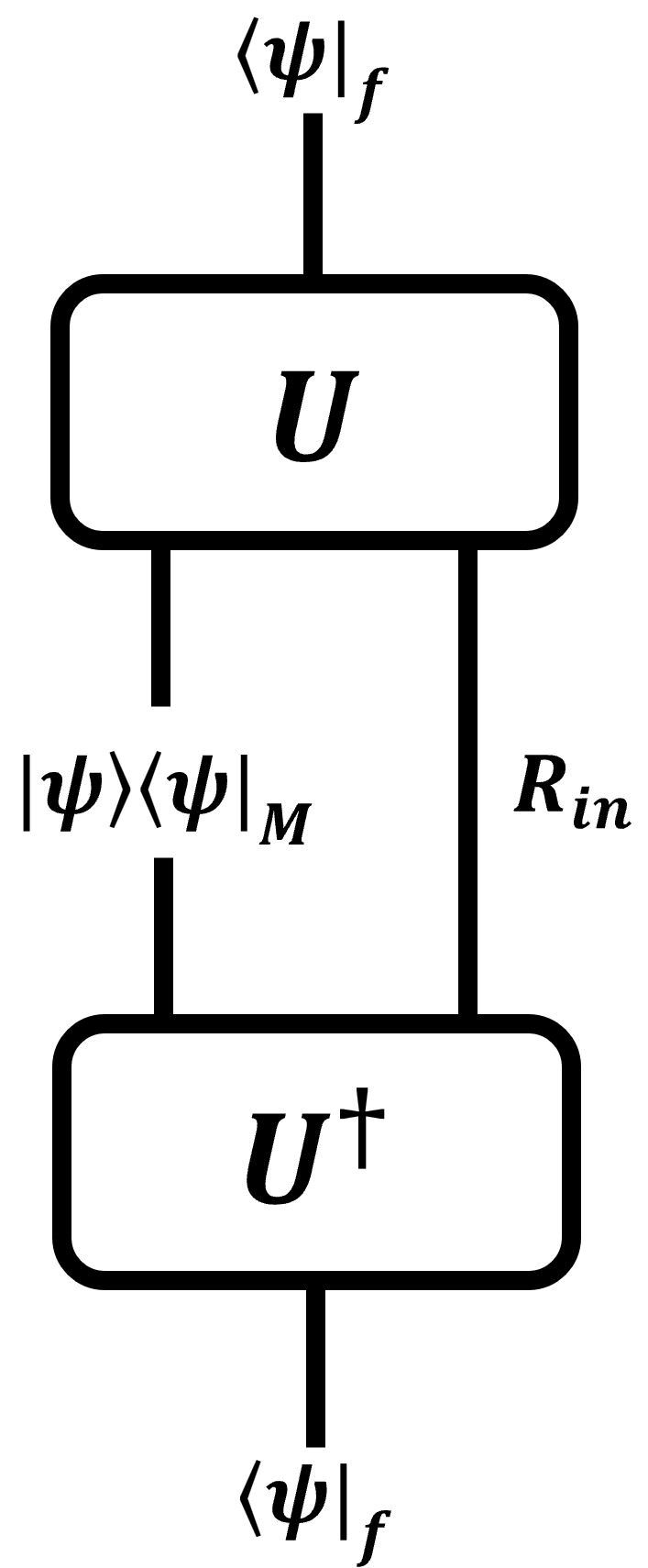}\right)\;.
\end{eqnarray}
It gives the same graphical representation of $\textrm{Tr}\rho_\Phi$ which is given in Eq.\eqref{Tr_rho_phi_graph}. Therefore, it can be checked that $|\Psi_{in}\rangle$ is also normalized under the Haar average
    \begin{eqnarray}
        \int dU \textrm{Tr}\left[|\Psi_{in}\rangle\langle\Psi_{in}|\right]=1\;.
    \end{eqnarray}

Using the graphical representation, the decoding probability is given by
\begin{eqnarray}
    P&=&\textrm{Tr}\left(|\psi\rangle_f\langle\psi|\Psi_{in}\rangle\langle\Psi_{in}| \right)\nonumber\\    
    &=& \quad\figbox{0.3}{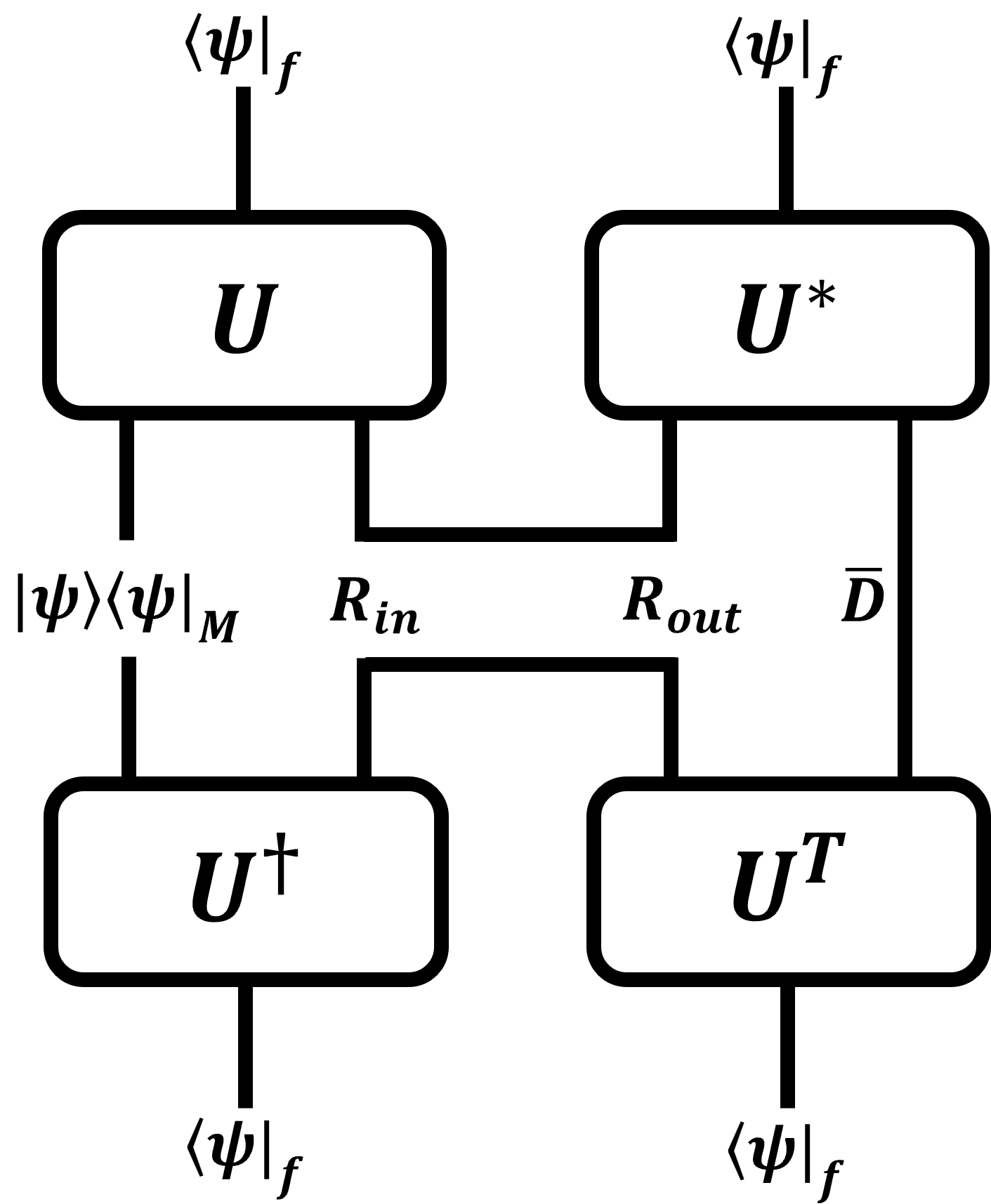}\;.
\end{eqnarray}
In order to calculate the Haar average of the decoding probability, the following Haar integral formula should be useful \cite{Kim:2022pfp,Li:2024tcm}
\begin{eqnarray}
       && \int dU\  U_{i_1j_1}U_{i_2j_2}U^\dagger_{j_1'i_1'}U^\dagger_{j_2'i_2'}\\
&=&\frac{1}{(d^2-1)}\left(\delta_{i_1i_1'}\delta_{i_2i_2'}\delta_{j_1j_1'}\delta_{j_2j_2'} + \delta_{i_1i_2'}\delta_{i_2i_1'}\delta_{j_1j_2'}\delta_{j_2j_1'}\right)
 \nonumber\\
 &-&\frac{1}{d(d^2-1)}\left(\delta_{i_1i_1'}\delta_{i_2i_2'}\delta_{j_1j_2'}\delta_{j_2j_1'} + \delta_{i_1i_2'}\delta_{i_2i_1'}\delta_{j_1j_1'}\delta_{j_2j_2'}\right)\,\nonumber\label{4U_integral}.
\end{eqnarray}
It is convenient to represent the integral formula in the graphical form 
\begin{eqnarray}
     &&\int dU \left(\figbox{0.25}{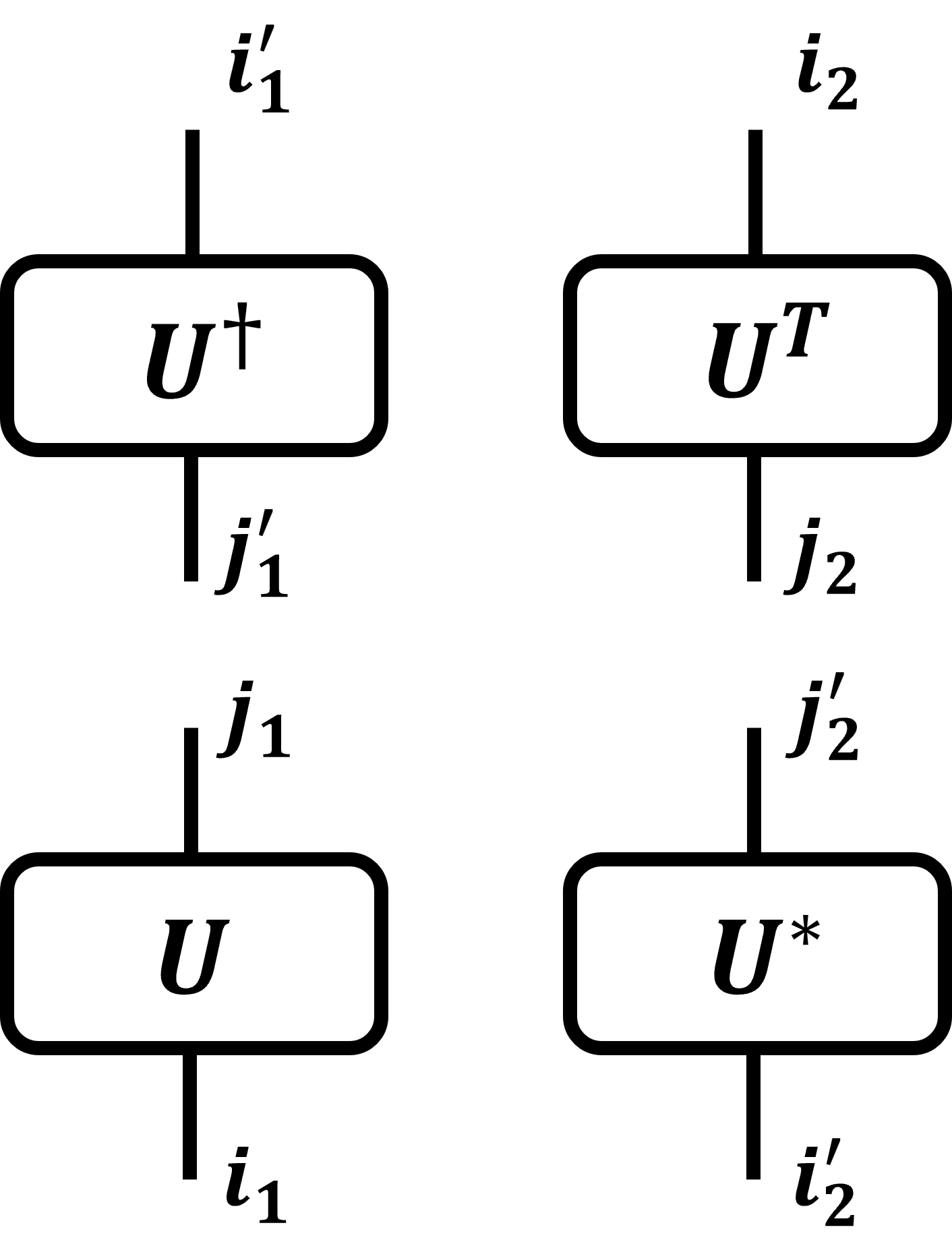}\right)
     \nonumber\\&=&\frac{1}{(d^2-1)}\left(\figbox{0.25}{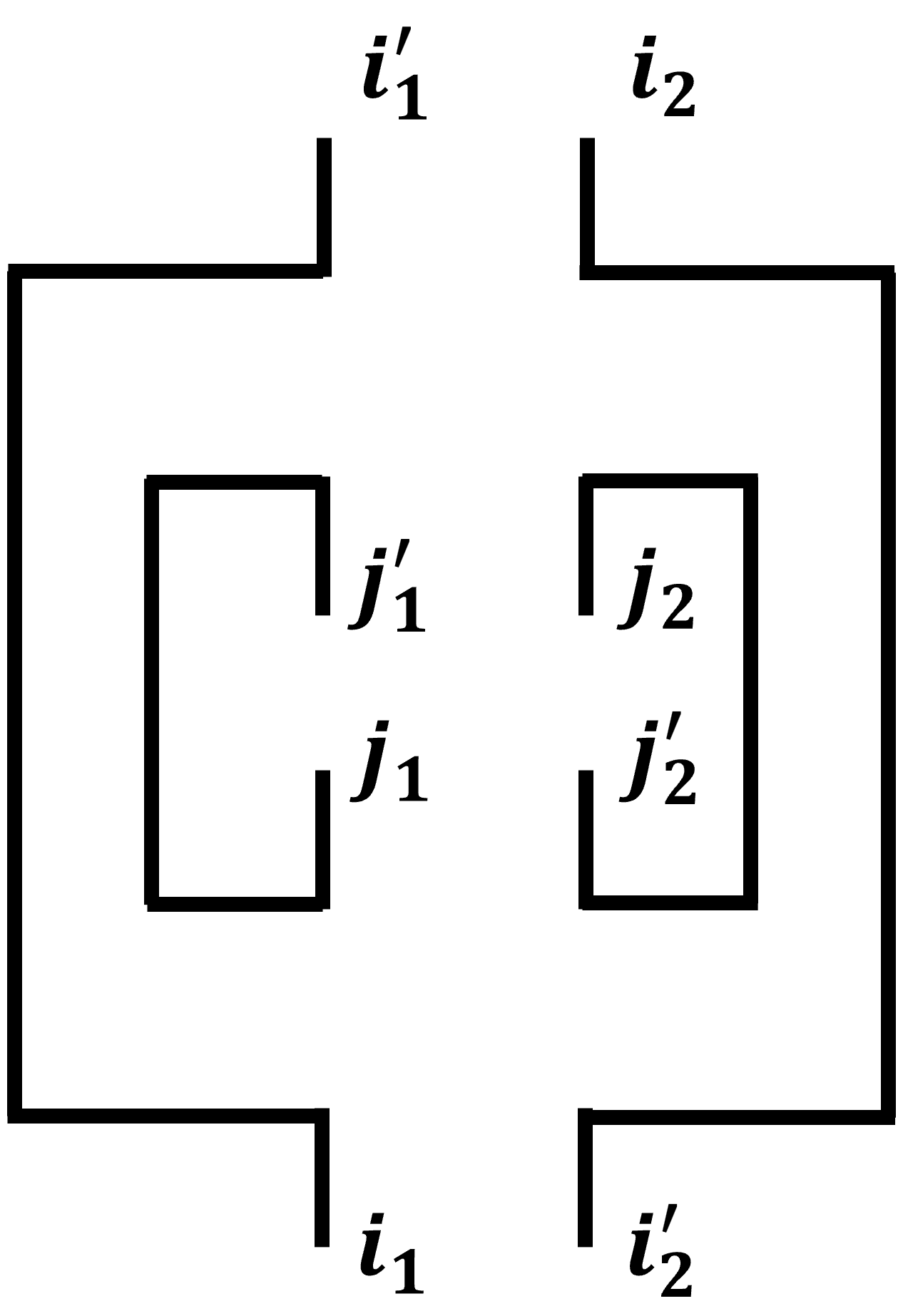}+\figbox{0.25}{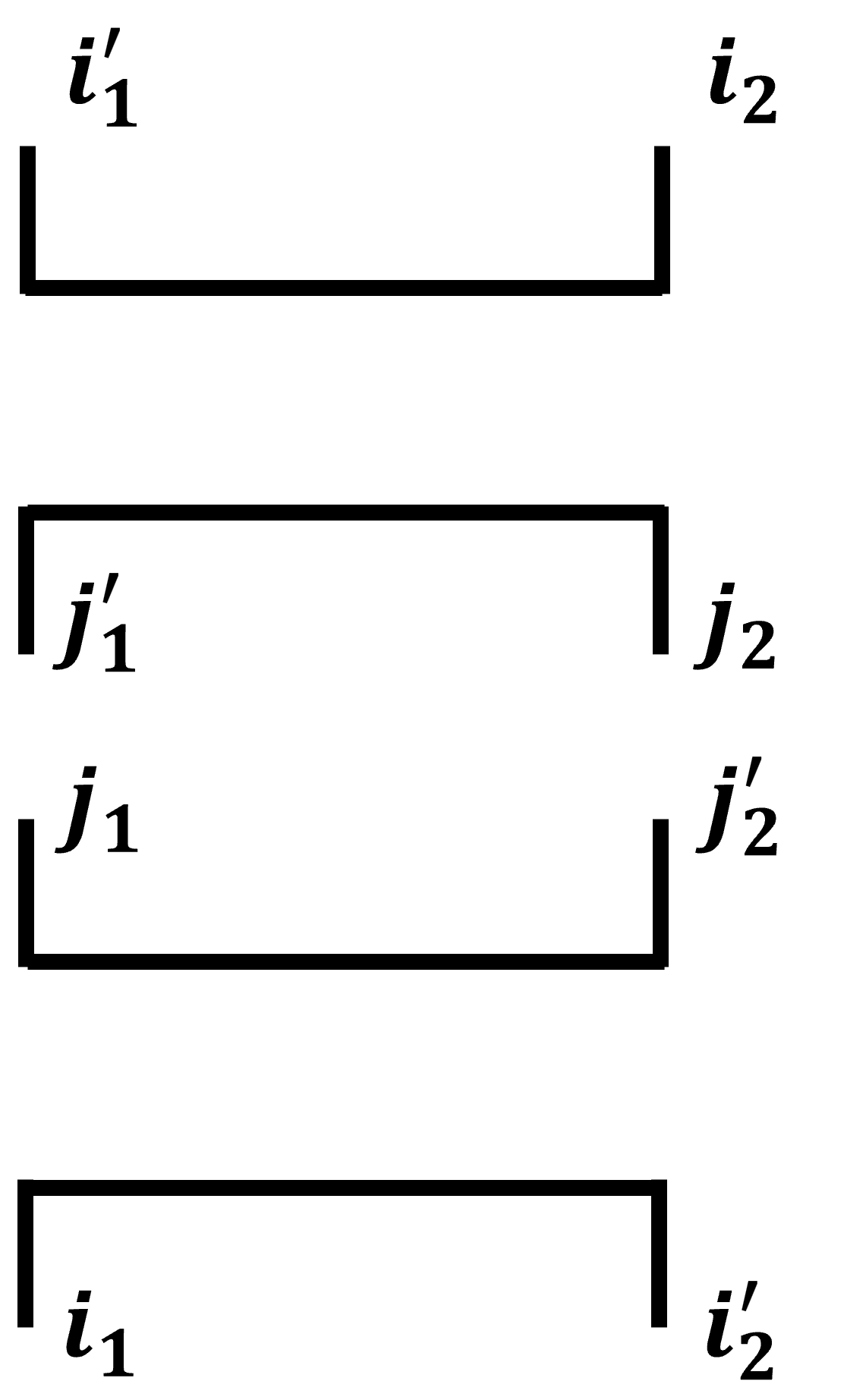}\right)\nonumber\\
    &+&\frac{1}{d(d^2-1)} \left(\figbox{0.25}{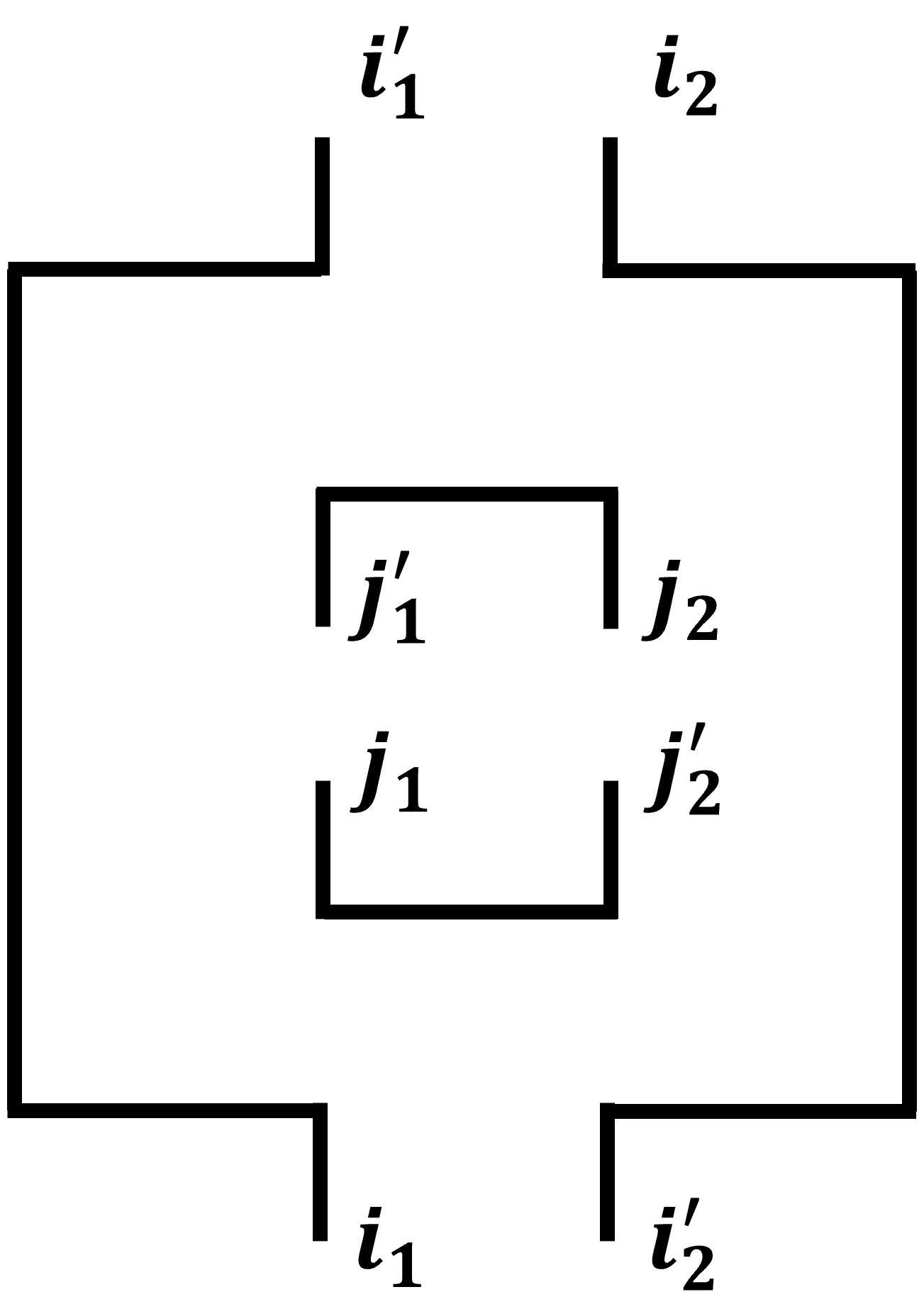}+\figbox{0.25}{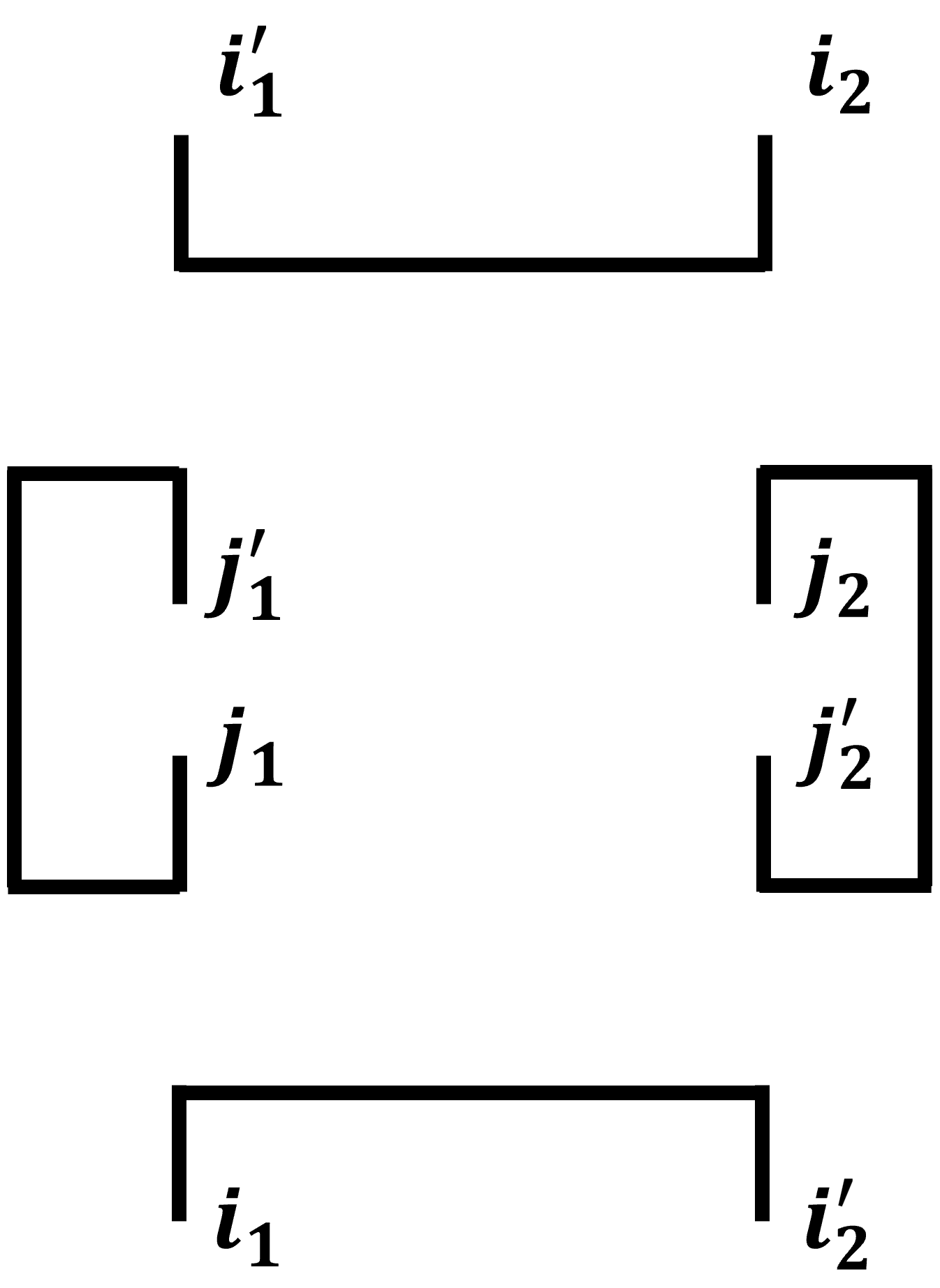}\right)\;.\label{4U_graph}
\end{eqnarray}

The Haar average of the decoding probability is then given by
\begin{eqnarray}\label{Decoding_Prob}
    \overline{P}&=&\int dU P\nonumber\\
    &=&\frac{1}{N^4-1}\left[ \left(\figbox{0.3}{probability1.png}\right)+\left(\figbox{0.3}{probability2.png}\right)\right]\nonumber\\
   &&-\frac{1}{N^2(N^4-1)} \left[\left(\figbox{0.3}{probability3.png}\right)+\left(\figbox{0.3}{probability4.png}\right)\right]
   \nonumber\\
  &=&\frac{2}{N^2+1}\;.
\end{eqnarray}  
When $N$ is sufficiently large, the average decoding probability is $\frac{2}{N^2}$, indicating an extremely small likelihood of successfully decoding the information from the radiation.

The decoding fidelity measured by the matching of the state $|\Psi_{out}\rangle$ with the initial state $|\psi\rangle_M$ for the matter system, which quantifies the quality of the decoding. The decoding fidelity can be calculated as 
\begin{eqnarray}\label{Decoding_Fid}
    F&=&\textrm{Tr}\left(|\psi\rangle_M\langle\psi|\Psi_{out}\rangle\langle\Psi_{out}| \right)\nonumber\\
    &=&\frac{N^2}{P} \quad \figbox{0.3}{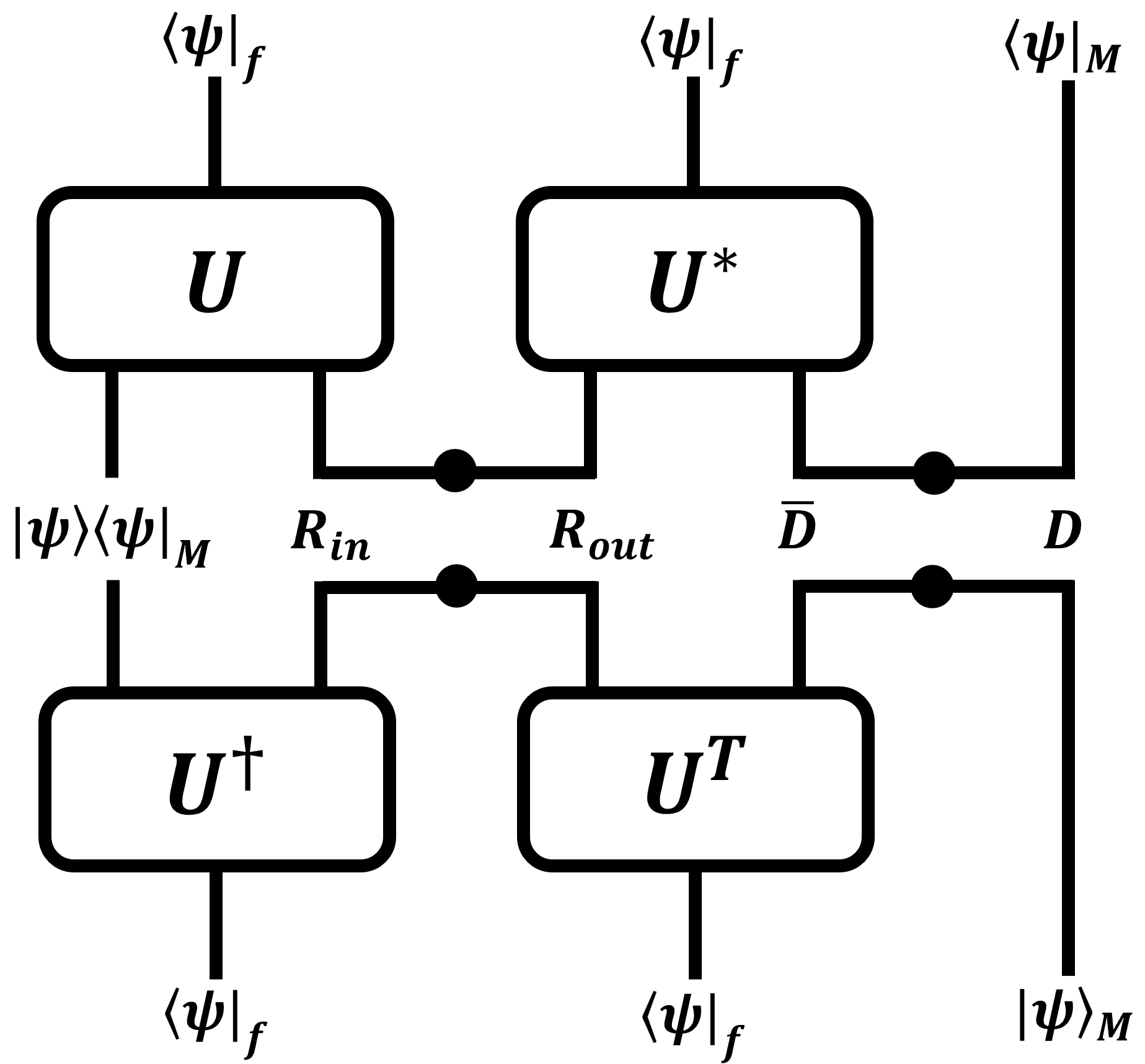}\nonumber\\
    &=&\frac{1}{P} \quad \figbox{0.3}{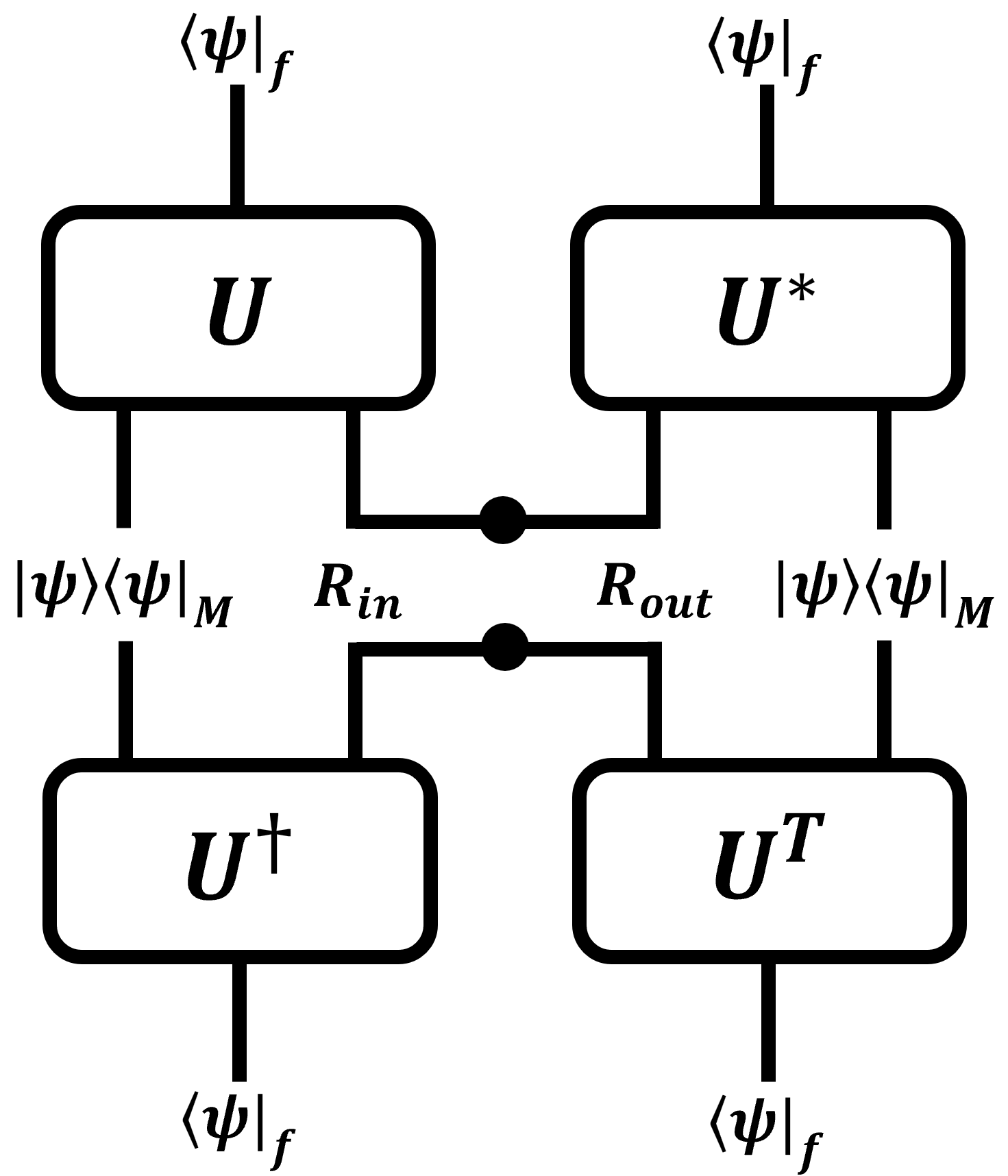}\;.
\end{eqnarray}

In average, one finds that the product of the decoding probability and the fidelity is given by 
\begin{eqnarray}
    \int dU PF&=&\frac{1}{N^4-1}\left[ \left(\figbox{0.3}{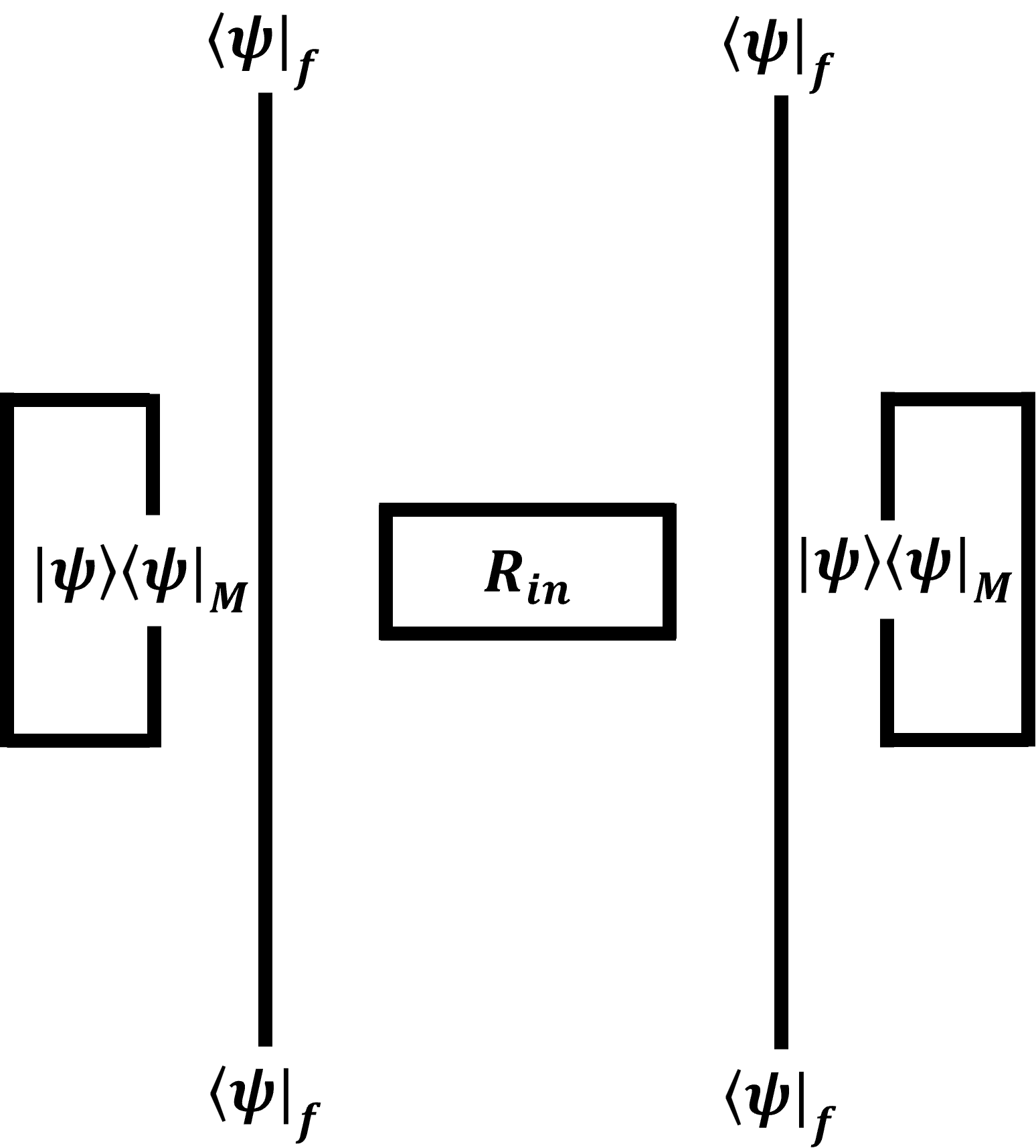}\right)+\left(\figbox{0.3}{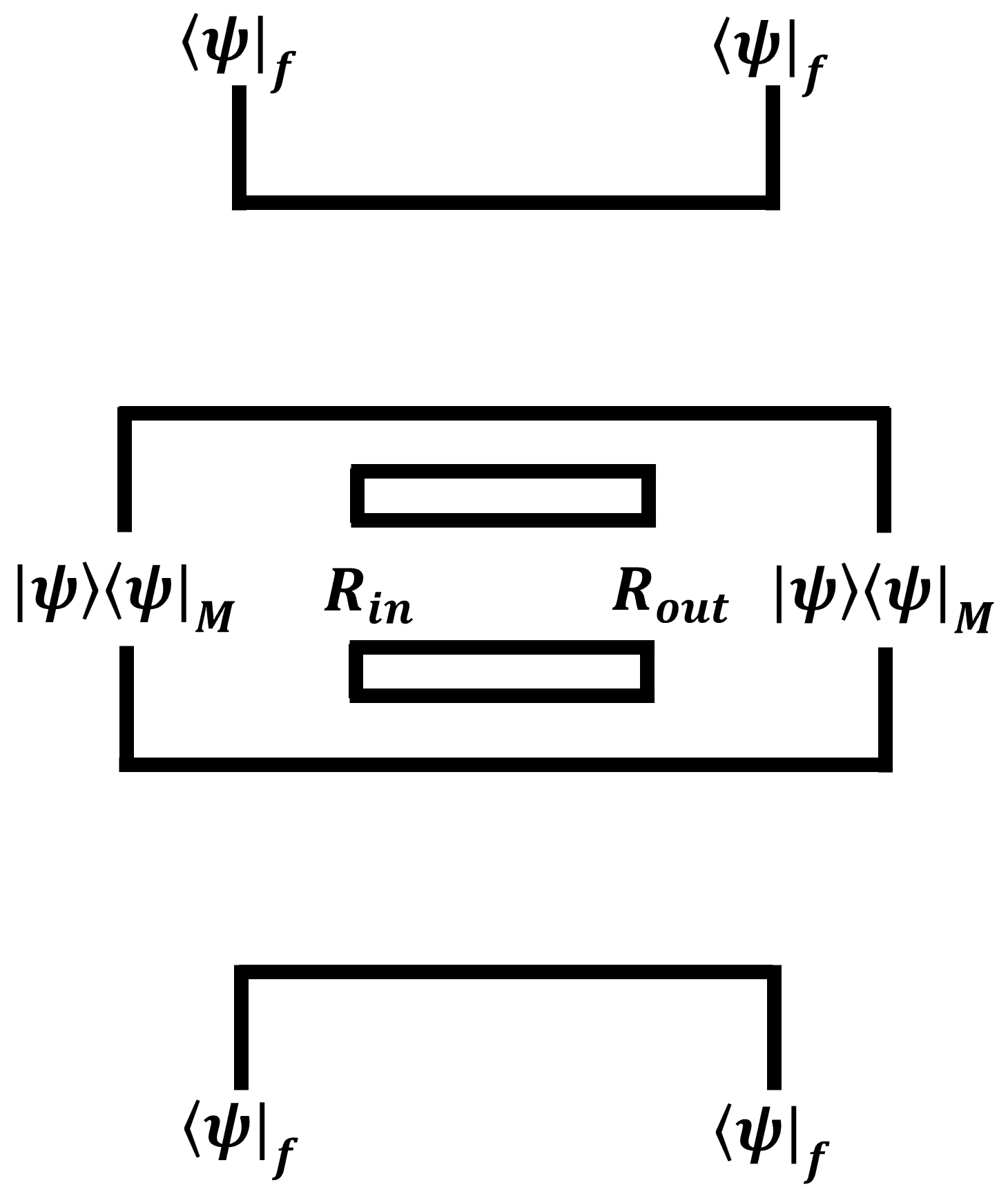}\right)\right]\nonumber\\
   &&-\frac{1}{N^2(N^4-1)} \left[\left(\figbox{0.3}{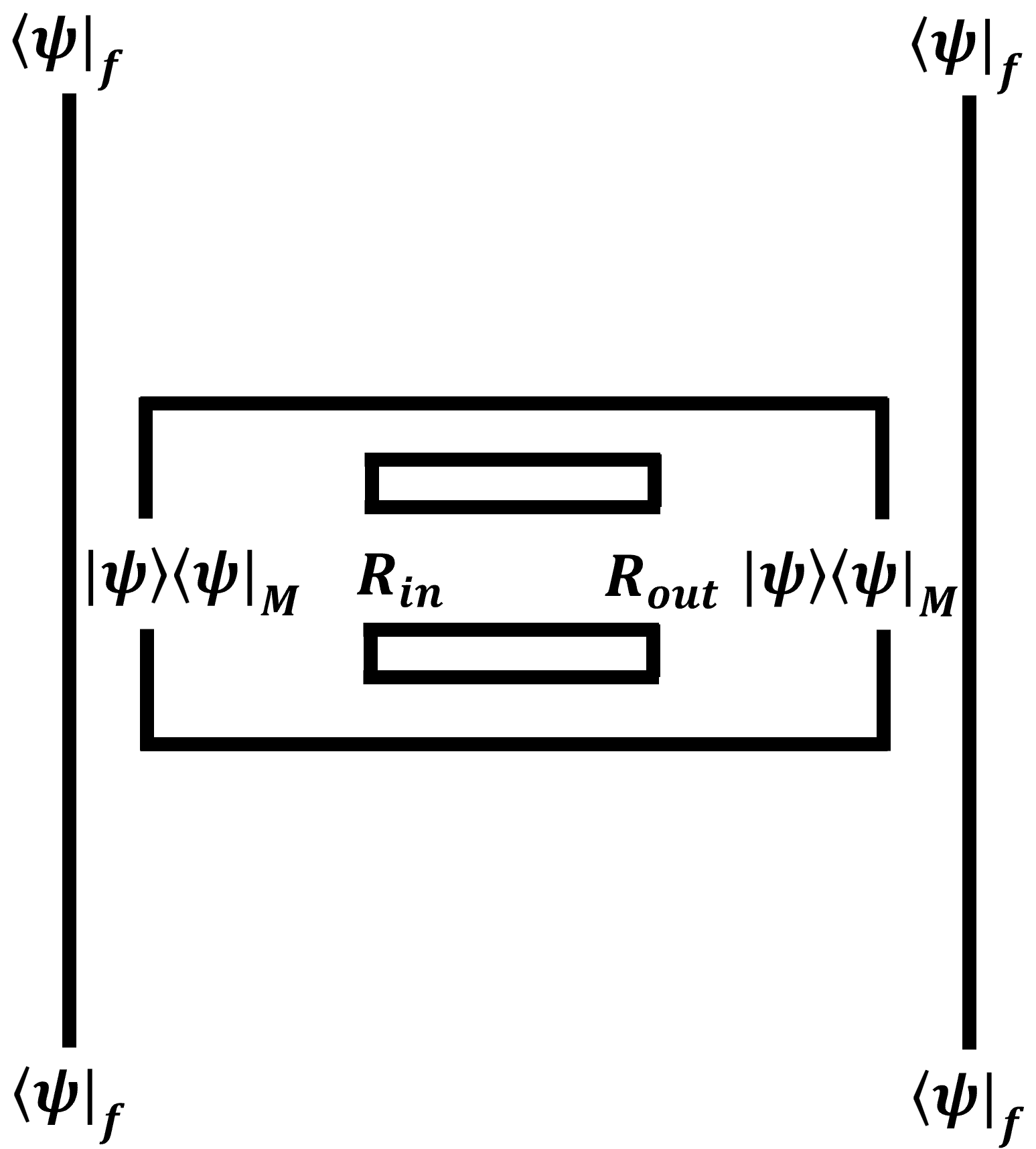}\right)+\left(\figbox{0.3}{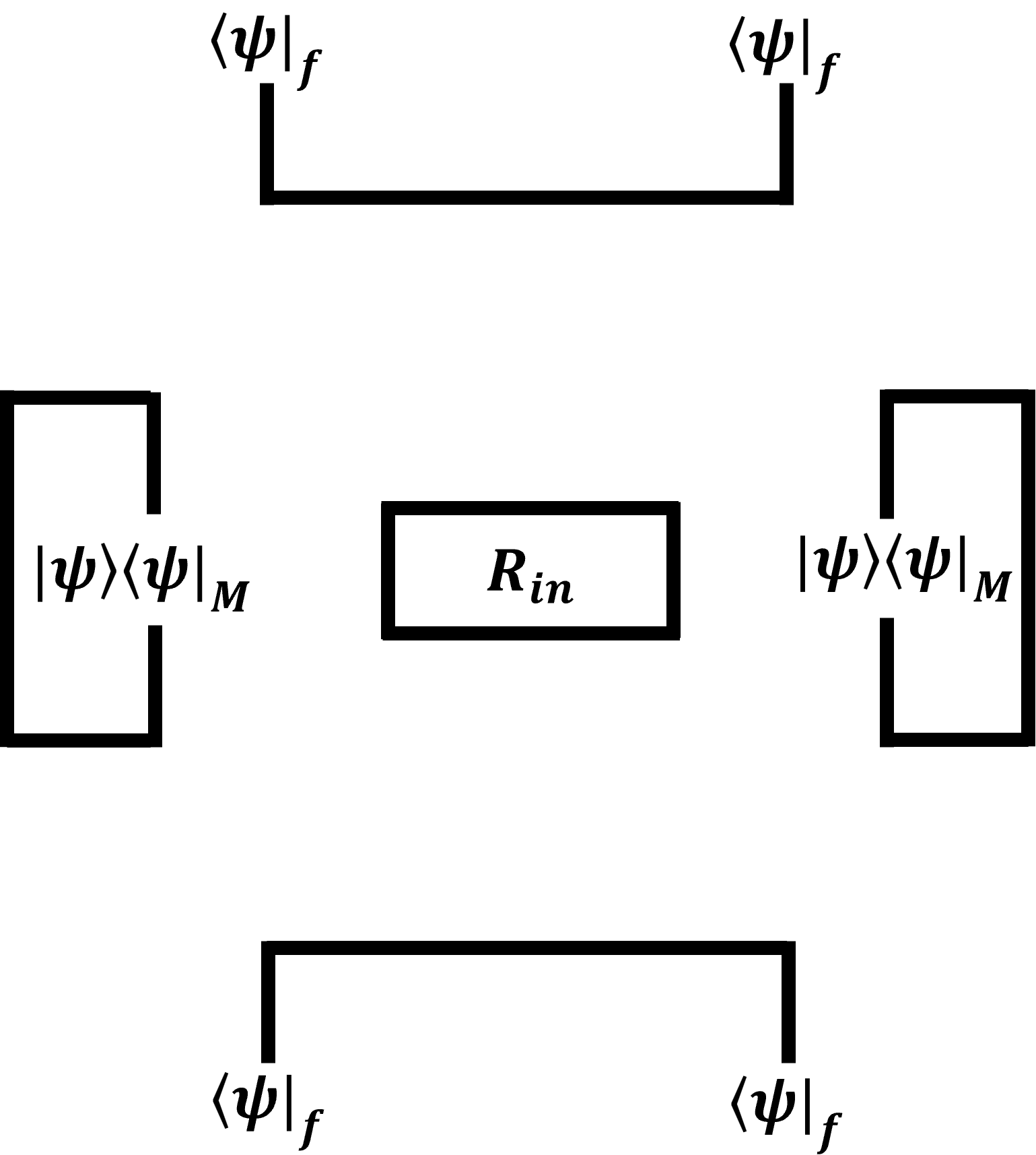}\right)\right]
   \nonumber\\&=&\frac{(N+1)}{N(N^2+1)}\;.
\end{eqnarray}
In the large $N$ limit, with the extremely small decoding probability $\frac{2}{N^2}$, the average fidelity can be estimated as $\overline{F}=\frac{1}{2}$ . It indicates that in the cost of extremely small decoding probability, the fidelity can be promoted to a relatively large value. However, this value can not reach unity. In conclusion, if considering the interactions between the collapsing matter system with the infalling interior Hawking partners, which is represented by an scrambling unitary operator, one cannot recover the information contained in the initial matter system completely.

\section{Conclusion}

In summary, our investigation focused on the fidelity of information retrieval within the black hole final state model, considering interactions between the collapsing matter and the infalling Hawking radiation within the event horizon. Utilizing a scrambling unitary operator to model these interactions, direct computation of the average fidelity suggests significant information loss during black hole evaporation. Subsequently, we applied the Yoshida-Kitaev decoding strategy for information recovery. While an improvement in decoding fidelity was observed, it remained finite and failed to reach unity. It should be noted that, due to the complicated dynamics within the black hole interior, the interactions between the collapsing matter and the infalling Hawking radiation are, in general, inevitable \cite{Sekino:2008he}. This suggests that the issue of unitarity in black hole evaporation may not be fully addressed within the framework of the final state model.

%\acknowledgments

\bibliographystyle{unsrt}
\bibliography{biblio.bib}
\end{document}